\documentclass[%
 reprint,
 amsmath,amssymb,
 aps,
]{revtex4-2}

\usepackage{graphicx}
\usepackage{dcolumn}
\usepackage{bm}
\usepackage{siunitx}

\begin{document}

\preprint{APS/123-QED}

\title{Constant speed penetration into granular materials:\\drag forces from the quasistatic to inertial regime}

\author{Leah K. Roth}%
 \email{rothl@uchicago.edu}
\affiliation{James Franck Institute, The University of Chicago, Chicago, Illinois 60637, USA}




\begin{abstract}
Predicting the force exerted on an object as it penetrates a granular medium is of interest in engineering, locomotive, and geotechnical applications.
Current models of granular drag, however, vary widely in applicability and parameterization, and the physical origin of the granular resistive force itself is a subject of debate.
Here we perform constant speed penetration experiments, combined with calibrated, large-scale molecular dynamics simulations, at velocities up to 2 m/s to test the effect of impact velocity on the depth dependent `hydrostatic' drag force.
We discover that the evolution of the granular flow field around an intruder regulates the presence of depth dependent drag forces.
In addition, we find that the observed linear depth dependence is commensurate with the mass of flowing grains. 
These results suggest that, as the impact speed increases beyond the quasistatic regime, the depth dependent drag term becomes intertwined with inertial effects.
\end{abstract}

\maketitle


\section{\label{sec:intro}Introduction}

Granular material occupies a unique position in engineering, robotics and geotechnical applications, due in large measure to its ability to behave as both a fluid and a solid.
At the same time, it is this very duality that proves an obstacle to a comprehensive understanding of granular impact~\cite{chandrasekar_kin_flow_2015,umbanhowar_scaling}.
Despite steady interest in analytically domesticating the varied responses of granular matter to impact and deformation, the prevailing model describing the drag force applied by a granular material on a penetrating object is phenomenological in origin, arising from studies of vertical projectile impact~\cite{goldman_gran_imp_2010}.

The drag force on an object intruding into a granular material a vertical penetration distance $z$ at velocity $v$ is almost universally separated into depth and velocity dependent terms, given by~\cite{goldman_gran_imp_2010,umbanhowar_scaling,goldman_robo,durian_unified_force,blumenfeld_soft_matter}

\begin{equation}\label{eq:f_drag}
    F_{\mathrm{drag}} = F_z\left(z\right)+F_v\left(z,v\right).
\end{equation}

The velocity dependent term $F_v$, thought to be an inertial force due to momentum transfer at higher impact speeds, has been shown to be independent of depth and quadratic with velocity in projectile experiments, so that $F_v = C \rho_g D^2 v^2$, where $\rho_g$ is the material density of the substrate, $D$ is the diameter of the intruder, and $C$ is a fitting constant~\cite{durian_unified_force,behringer_particle_scale,goldman_robo}.
$F_v$ is only effective above a critical impact velocity given by the speed of an individual grain settling under gravity $v_c = \sqrt{2gd_g}$, for gravity $g$ and grain diameter $d_g$~\cite{blumenfeld_nature,schiffer_vc}.
For $v < v_c$ the system is in the quasistatic regime, while at higher impact speeds the grains are predicted to be fluidized~\cite{blumenfeld_nature,schiffer_vc}.

The depth dependent term has been found to be linear in depth, so that $F_z = k_z z$~\cite{blumenfeld_soft_matter,blumenfeld_nature}.
$F_z$ emerges from frictional interactions between grains in the bed during impact, and varies with the gravitational loading of granular contacts~\cite{goldman_robo,durian_depth_dependent}.
Brzinski III \textit{et al.} have found $\text{d}\textbf{F}_z \sim \mu \left(\rho_g g z\right) \text{d}\textbf{A}$, where $\mu$ is equal to the tangent of the angle of repose and $\text{d}\textbf{A}$ is an area element normal to the projectile surface~\cite{durian_depth_dependent,durian_unified_force}.
For a cylinder, then, $F_z = \alpha \mu \rho_g g D^2 z$, where $\alpha$ is a constant that Brzinski III \textit{et al.} found to be $\sim 20-30$ across all experiments conducted~\cite{durian_depth_dependent,durian_unified_force}.
Though $F_z$ is due to forces that act normal to the projectile surface, the inter-grain friction features prominently, indicating that the motion of grains tangential to the intruder surface do not significantly contribute to the drag force and also that frictional grain-grain interactions must be engaged elsewhere during impact~\cite{durian_depth_dependent}.
In addition, a purely ``Archimedean'' force, directly proportional to the displacement of grains due to the penetration of an intruder, is an insignificant fraction of the observed drag force --- manifest in the large coefficient measured by Brzinski III \textit{et al.}~\cite{gondret_exp_vel,melo_dynamics_shear,durian_depth_dependent}.

Despite its widespread acceptance and use, the physical origins of this model, as well as specific functional forms and parameterizations, are still a matter of debate~\cite{durian_depth_dependent,blumenfeld_nature}.
Brzinski III \textit{et al.} suggest that the surprisingly large coefficient of the linear depth dependence is due to rigid-body motion of force chains that stretch into the bulk, which are mobilized during contact with the intruder surface and whose strength is mediated via gravitational loading~\cite{durian_depth_dependent}.
Alternatively, Kang \textit{et al.}, in their theoretical analysis of a modified Archimedes' law for dry grains, propose that a jammed stagnant zone preceding the intruder allows the intruder to interact with and dislodge a large volume of grains whose mass accounts for the strength of the resulting depth dependence~\cite{blumenfeld_nature}.
The size of this interaction region is then determined solely by the internal friction angle~\cite{blumenfeld_nature}.

In addition to the physical pedigree of the granular drag depth dependence, the coupling between low velocity effects and the velocity dependent force is not clear --- in particular, whether $F_{\mathrm{drag}}$ is truly additive, as in Eq.~\ref{eq:f_drag}.
The majority of granular drag experiments that probe the velocity dependent force have been performed by dropping a projecile into a bed of grains, which results in strong coupling between the intruder velocity and penetration depth and which may make it difficult to completely separate the components of $F_{\mathrm{drag}}$ or to determine if there are indeed two independent components.

We present granular drag experiments and simulations conducted at a constant velocity, both near and above $v_c$, in an effort to systematically characterize the relationship between $F_v$ and $F_z$.
We perform constant speed vertical penetration experiments over four orders of magnitude in velocity and use these results to calibrate molecular dynamics simulations that enable us to directly characterize the grain-scale effects in three dimensions.

\section{\label{sec:background}Background}

The most recent experimental investigation of the depth dependent drag force was performed by Kang \textit{et al.}~\cite{blumenfeld_nature}.
In these experiments, a cylindrical intruder is pushed into a granular bed at a constant velocity much less than the quasistatic limit $v \ll v_c$~\cite{blumenfeld_nature}.
Kang \textit{et al.} find that the linear force versus depth curves for a wide range of materials and intruder shapes can be collapsed with the use of a parameter, $K_{\phi}$, which is a function of the internal angle of friction~\cite{blumenfeld_nature}.
However, Kang \textit{et al.} also observe a transient nonlinear regime at shallow penetration depths, preceding the steady-state, linear drag force regime~\cite{blumenfeld_soft_matter,blumenfeld_nature}.
In a subsequent publication, Feng \textit{et al.} identify three regimes~\cite{blumenfeld_soft_matter}.
The very early stages of impact are characterized by compression of the bed, which in turn begins to form a `stagnant zone' (SZ) immediately ahead of the intruder~\cite{blumenfeld_soft_matter}.
An intermediate regime follows this compression, characterized by yielding and plastic flow, during which the SZ is steadily growing in size~\cite{blumenfeld_soft_matter}.
Once the SZ has developed fully, characterized by a dense, conical structure underneath the intruder, the linear depth dependent force regime commences~\cite{blumenfeld_soft_matter}.
During penetration, the SZ will effectively cleave the undisturbed material and push it to the sides~\cite{blumenfeld_soft_matter}.
In numerical studies by Feng \textit{et al.}, the internal stress field in the steady state regime in simulation is used to derive an effective internal friction angle, while $K_{\phi}$ is fitted from the depth dependent force, yielding results consistent with the experiments performed in Kang \textit{et al.}~\cite{blumenfeld_soft_matter}.

Experiments by Aguilar and Goldman have investigated the jump height of a robotic mass-spring system off of a granular substrate, including a dependence on packing fraction and wait time between successive jumps~\cite{goldman_robo}.
To explain discrepancies in predictions of jump height in situations of repeated push-off, Aguilar and Goldman develop a model of granular impact consisting of a ``jammed granular cone'' that increases in size and mass as a function of intruder depth~\cite{goldman_robo}.
This is similar to Kang \textit{et al.}'s description of a conical stagnant zone that grows beneath the intruding object, though in this case it is extended to describe the behavior of impacts above $v_c$ under dynamic conditions~\cite{blumenfeld_nature,goldman_robo}.
In addition, Aguilar and Goldman visualized the development of the added mass cone by performing a robotic jump test near a transparent wall, tracking the positions of the grains during impact, and calculating the subsequent shear strain rate $\dot{\gamma}$~\cite{goldman_robo}.
Using this method, Aguilar and Goldman observed shear bands extending downwards in a triangular fashion from the edges of the intruding surface, which they interpreted as designating the boundaries of a jammed cone that grows, and then proceeds, ahead of the intruder and ``wedges'' the material to the sides~\cite{goldman_robo}.

Kang \textit{et al.} claim that ``several experiments [have] established that the penetration process gives rise to a conical SZ ahead of the intruder, advancing as a rigid body''~\cite{blumenfeld_nature}.
However, many of the experiments cited as support for the conical SZ are performed in a quasiplanar geometry or near a transparent boundary, which could significantly distort the flow field in ways that would seem to increase the likelihood of generating a stagnant zone beneath a flat intruder~\cite{melo_effect_of_pf,melo_dynamics_shear,chandra_deformation_field,chandrasekar_kin_flow_2015,goldman_robo}.
Significantly, experiments performed by Hamm \textit{et al.} and Tapia \textit{et al.} with a semicircular intruder in a quasiplanar geometry instead observe \textit{dilation} in the dual shear bands that are caused by the penetration and extend off of either side of the intruder~\cite{melo_dynamics_shear,melo_effect_of_pf}.
They do not appear to observe a conical (or triangular) jammed region moving at the speed of the intruder in this case, but rather symmetric shear bands that extend outwards from the bottom edge of intruder towards the surface of the bed~\cite{melo_dynamics_shear}.

This is similar in character and structure to the shear bands observed in horizontal plate drag by Gravish \textit{et al.} and Kobayakawa \textit{et al.}~\cite{goldman_force_flow_prl,goldman_force_flow_pre,tanaka_plate_drag}.
The drag force exerted on the plate can be divided into two regimes: an initial, transient drag force during yielding, and a steady state drag that remains constant, on average~\cite{goldman_force_flow_pre}.
Though the mean steady state drag force exerted on the plate increases linearly with the packing fraction of the bed, the emergence of temporal fluctuations riding on top of the mean force mark the critical packing fraction, $\phi_c$, at which dilation begins ~\cite{goldman_force_flow_prl}.
Above $\phi_c$, in this case 60.3\%, stationary shear bands are created ahead of the plate~\cite{goldman_force_flow_prl}.
The shear bands focus dilation and deformation by weakening the material, and they correlate with fluctuations in the drag force~\cite{goldman_force_flow_prl}.
The onset of dilation, in which the packing fraction of the bed surpasses $\phi_c$, is also marked by a peak in the yield force~\cite{goldman_force_flow_pre}.
Significantly, the height of this force peak depends on the velocity of the plate, though the position of the peak remains independent of velocity~\cite{goldman_force_flow_pre}.
The end of the peak regime corresponds to the development of a stable shear band, so that flow only exists between the shear band and the plate~\cite{goldman_force_flow_pre}.

Discrete element method (DEM) simulations of horizontal plate drag by Kobayakawa \textit{et al.} confirm the results of Gravish \textit{et al.}~\cite{tanaka_plate_drag}.
Though the mean force increases linearly with the packing fraction of the bed $\phi_0$, periodic oscillations in the drag force emerge when $\phi_0 > \phi_c$ due to the creation and destruction of stationary shear bands~\cite{tanaka_plate_drag}.
Significantly, when $\phi_0 > \phi_c$, the packing fraction behind the shear band and ahead of the plate is given by $\phi_s$, which is determined by the specific flow conditions and is actually lower than $\phi_c$~\cite{tanaka_plate_drag}.
This dilated region is ``strongly confined'' due to the difference in stability between the flowing and static grains~\cite{tanaka_plate_drag}.
However, Kobayakawa \textit{et al.} found that larger grains relative to the intruder dimension destroy the periodicity of the force fluctuations, indicating that relatively larger grains hamper the development of distinct shear bands~\cite{tanaka_plate_drag}.

Seguin \textit{et al.} performed experiments of a horizontal cylinder penetrating a bed of glass beads in a quasiplanar geometry at velocities from 0.5 to 50 mm/s, so as to remain below the quasistatic limit $v_c$~\cite{gondret_exp_vel}.
The initial packing fraction of 0.62 ensured that the system was ``dense,'' or above $\phi_c$~\cite{gondret_exp_vel}.
Seguin \textit{et al.} find that the radial and azimuthal velocity profiles do not evolve in time, indicating that the flow is in a steady state, though the force on the cylinder in increasing approximately linearly with penetration depth $z$~\cite{gondret_exp_vel}.
Both the radial and azimuthal flow is strongly localized around the cylinder which then falls off exponentially with distance, marking this as an area of high shear~\cite{gondret_exp_vel}.
The characteristic length of the flow region observed by Seguin \textit{et al.} is independent of the intruder velocity, and is independent of grain size as long as the ratio of intruder diameter to grain diameter is greater than $\sim 10$~\cite{gondret_exp_vel}.
Seguin \textit{et al.} likened the high granular temperature and high strain rates near the intruder to the temperature in a viscous fluid, where heat diffuses into the bulk and lowers the viscosity~\cite{gondret_exp_vel}.

\section{\label{sec:method}Methods}

\subsection{\label{sec:exp_meth}Experiment}

\begin{figure}[h]
\includegraphics[width=\columnwidth]{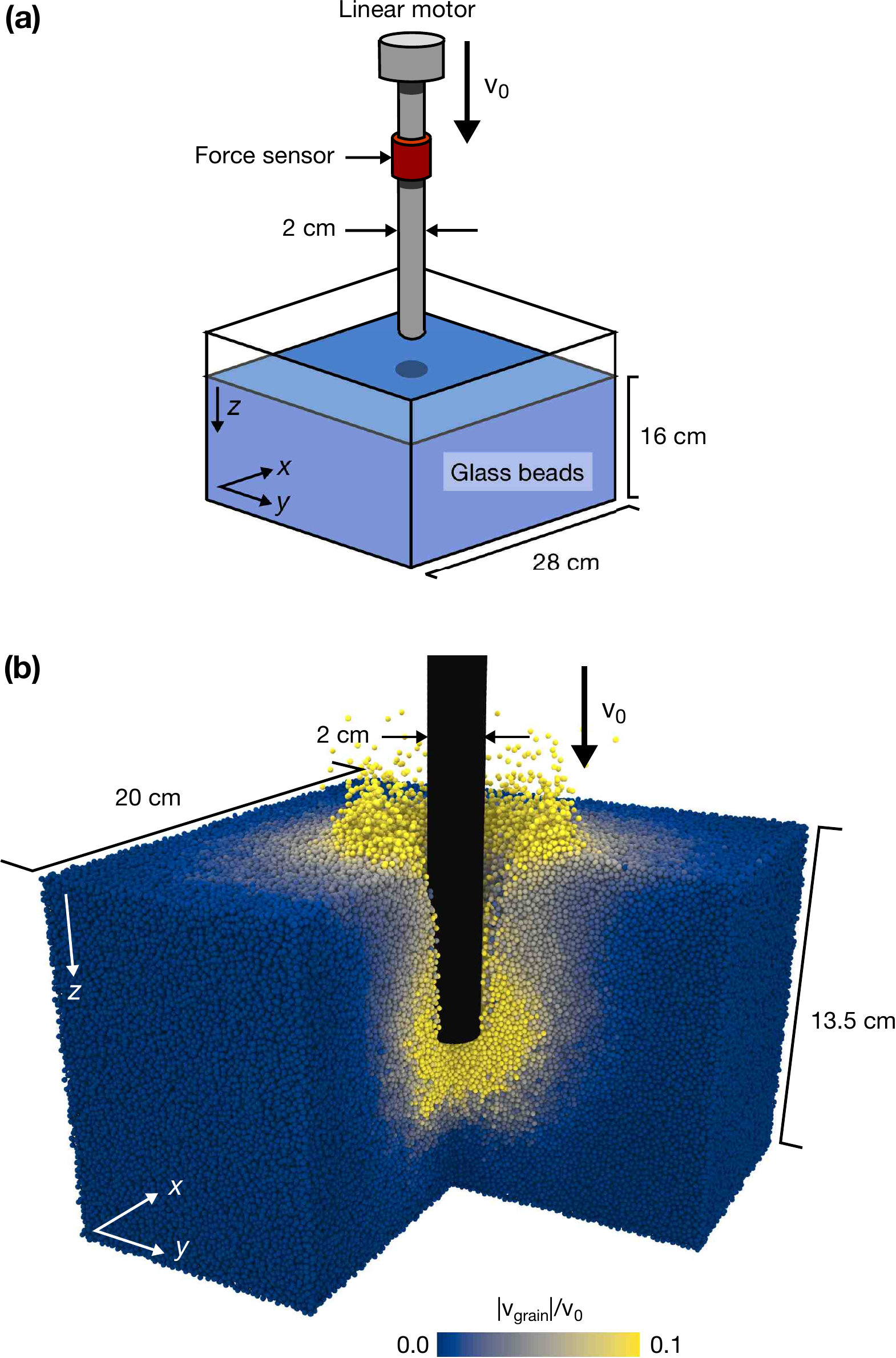}
\caption{\label{fig:setup_exp_sim} \textit{a)} Diagram of the experimental setup (not to scale). \textit{b)} Snapshot of the simulation setup. Grains are colored by velocity relative to the intruder speed $v_0=200$ cm/s.}
\end{figure}

The experimental setup is diagrammed in Fig.~\ref{fig:setup_exp_sim}, \textit{a}.
A rectangular box, with side length 28 cm, was filled to a depth of 16 cm with glass beads (Ceroglass beads, diameter 1.7---2.1 mm and density 2.8 $\pm$ 0.8 g/cm\textsuperscript{3}).
The bed was prepared by pouring the beads into the container and leveling by hand.
After every experimental run, the box was emptied and re-poured.
A steel rod with a diameter $D$ of 2.0 cm was attached to a Parker electric tubular linear motor and pushed into the bed at a constant velocity, with impact speeds $v_0$ ranging from 40 to 200 cm/s.
The maximum rod penetration depth was set to 10 cm to avoid effects from the bottom boundary.
A force sensor (Dytran Instruments 1051V2) in series with the intruding rod measured the total penetration resistance as a function of time with a sampling rate of 5-12 kHz, depending on the impact velocity.
These data, along with the impact velocity $v_0$, was used to produce the drag force acting on the rod as a function of penetration depth.

For low impact velocities, from 0.1 to 5.0 mm/s, the rod was coupled with an Instron 5800 series material tester, which measured both the drag force and penetration depth at a sampling rate of 550 Hz.
For each experimental drag force curve in what follows, five individual runs were averaged for $v_0 < 40$ cm/s and three individual runs were averaged for $v_0 \geq 40$ cm/s.

\subsection{\label{sec:sim_meth}Simulation}

We used the molecular dynamics simulation platform LAMMPS~\cite{lammps} to simulate constant speed penetration of a rod into a granular bed, shown in Fig.~\ref{fig:setup_exp_sim}, \textit{b}.
A container of side length 20 cm, periodic in $\hat{\mathbf{x}}$ and $\hat{\mathbf{y}}$, was filled to a height of $\sim 13.5$ cm with grains of diameter $d_g = 0.2$ cm.
A rod, 2.0 cm in diameter, was composed of $\sim 560$ overlapping spheres with identical pairwise interactions as those between individual grains.
The rod then penetrated the granular bed at a constant velocity from 40 to 240 cm/s to a depth of 10 cm.
The depth and width of the simulation box is 16\% and 28\% smaller than the experiment, respectively.
However, the simulation results do not change after shrinking the container dimensions or introducing sidewalls, indicating that boundary effects are not influencing the results.

The bed was prepared by generating a dilute, random configuration of grains inside the box, which was then allowed to fall towards the bottom of the container under gravity.
After the grains reached their minimum position, the kinetic energy of the system was periodically removed to prevent further densification through vibration.
The final configuration was not found to be sensitive to the relaxation timescale.
This preparation method is designed to approximate the pouring method used in the experiment, and produces a bed whose packing fraction is both uniform in depth and sensitive to the strength of frictional interactions between grains.

A frictional Hertzian contact force was used, where the normal force, $\mathbf{F}_{n_{ij}}$, and tangential force, $\mathbf{F}_{t_{ij}}$, on particle $i$ due to particle $j$ is given by

\begin{eqnarray}
\textbf{F}_{n_{ij}} = \sqrt{\delta_{ij}}R_{\mathit{eff}}\left(k_n \delta_{ij} \textbf{n}_{ij}-m_{\mathit{eff}} \gamma_n \textbf{v}_{n_{ij}}\right)
\\
\textbf{F}_{t_{ij}} = \sqrt{\delta_{ij}}R_{\mathit{eff}}\left(-k_t \textbf{u}_{n_{ij}}-m_{\mathit{eff}} \gamma_t \textbf{v}_{t_{ij}}\right)
\label{eq:hertz1}
\end{eqnarray}

Here, $R_{\mathit{eff}} = \left(R_i R_j/\left(R_i+R_j\right)\right)^{-1/2}$, $m_{\mathit{eff}}=m_im_j/\left(m_i+m_j\right)$, $R_i$ and $m_i$ are the radius and mass of particle $i$, $\mathbf{n}_{ij}$ is a unit vector along the centers of spheres $i$ and $j$, and $\mathbf{v}_{n_{ij}}$ and $\mathbf{v}_{t_{ij}}$ are the normal and tangential components, respectively, of the relative velocity between particles $i$ and $j$ ~\cite{silbert_hertz}.
The overlap between particles $i$ and $j$ is given by $\delta_{ij}$, $k_n$ and $k_t$ are the normal and tangential elastic constants, and $\gamma_n$ and $\gamma_t$ are the normal and tangential viscoelastic damping constants.
The rate of change of tangential displacement, $\mathbf{u}_{n_{ij}}$, is truncated to ensure $\left|\mathbf{F}_{t_{ij}}\right| < \left|\mu \mathbf{F}_{n_{ij}}\right|$, where $\mu$ is the coefficient of friction~\cite{silbert_hertz}.

\begin{table}[h]
\caption{\label{tab:sim_param}%
Simulation parameters.
}
\begin{ruledtabular}
\begin{tabular}{lcdr}
\textrm{Parameter}&
\textrm{Value}\\
\colrule
Rod diameter $D$ (cm) & 2.0\\
Particle diameter $d_g$ (cm) & 0.2\\
Particle density $\rho_g$ (g/cm\textsuperscript{3}) & 3.0\\
Normal stiffness $k_n$ (Pa) & \SI{1e9}{}\\
$k_n/k_t$ & 1\\
Normal damping $\gamma_n$ (s\textsuperscript{-1}cm\textsuperscript{-1}) & 1000\\
$\gamma_n/\gamma_t$ & 1\\
Time step $\delta t$ (s) & \SI{2e-7}{}\\
Friction coefficient $\mu$ & 0.3\\
\end{tabular}
\end{ruledtabular}
\end{table}

The simulation parameters were chosen to match the experiment as closely as possible, and are shown in Table~\ref{tab:sim_param}.
The normal stiffness, $k_n$ = \SI{1e9}{} Pa, is realistic for glass spheres of this size~\cite{silbert_hertz,silbert_mu}, and we performed simulations for $k_n$ = \SI{1e8}{} Pa and \SI{1e10}{} Pa with no change in results.
The ratio between the normal and tangential stiffness, $k_n/k_t = 1$, as well as the ratio between the damping constants, $\gamma_n/\gamma_t = 1$, have been chosen for simplicity, and do not change the results~\cite{silbert_hertz}.

\section{\label{sec:exp_results}Experimental Results}

\begin{figure}[h]
\includegraphics[width=\columnwidth]{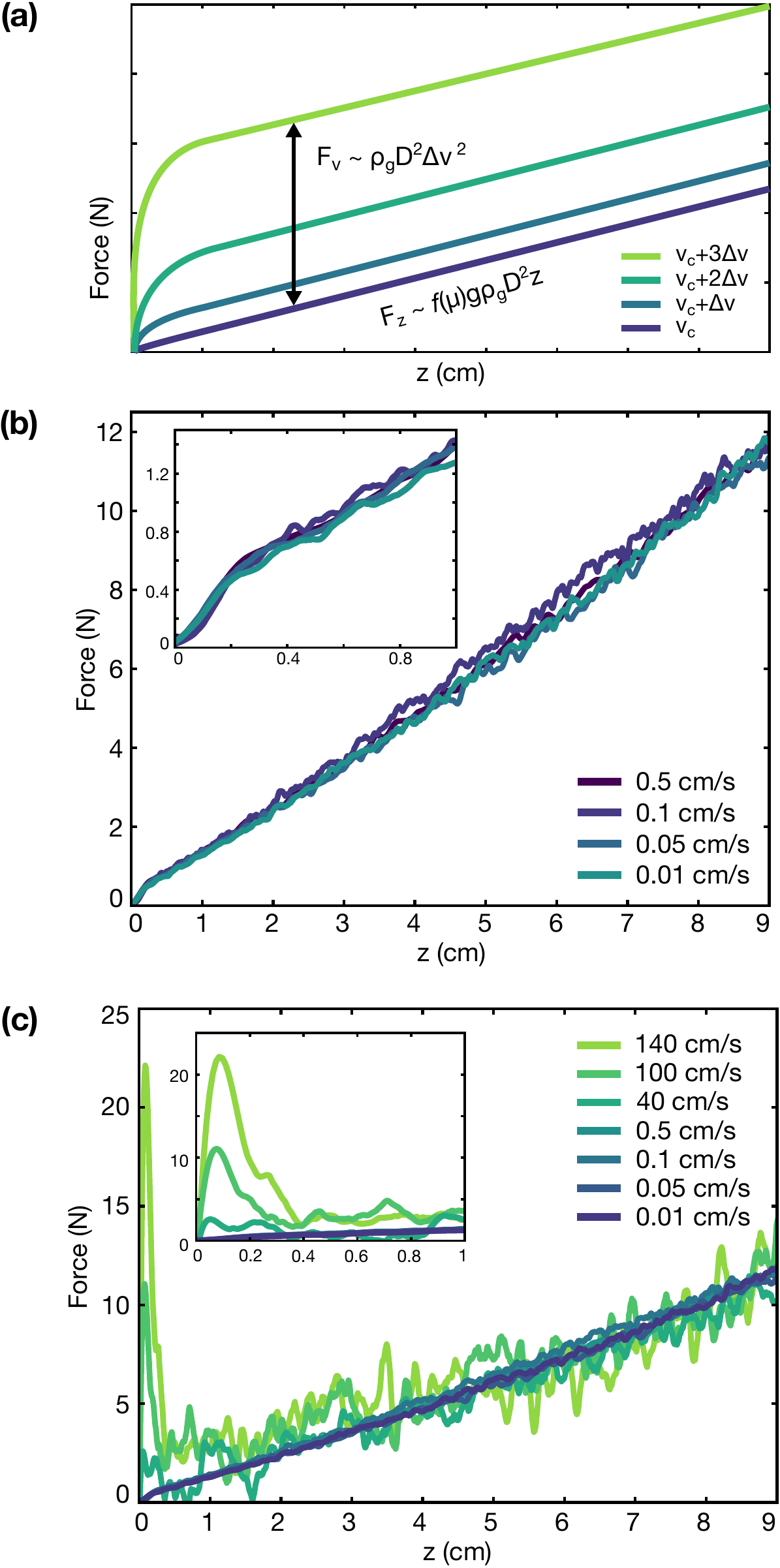}
\caption{\label{fig:exp_slow_fast} \textit{a)}  Diagram of the expected results, when extrapolating from existing models of impact above quasistatic velocities. \textit{b)} Experimental granular drag force felt by an intruder impacting at quasistatic velocities. \textit{c)} Experimental granular drag force felt by an intruder impacting above and below quasistatic velocities.}
\end{figure}

In order to confirm the results of previous quasistatic penetration experiments, the first set of experiments were performed at intruder velocities well below the quasistatic limit $v_c$~\cite{schiffer_vc,blumenfeld_nature}.
For the grains used here, with a diameter $d_g \approx 2$ mm, $v_c \sim 20$ cm/s, so these initial tests were confined to velocities below $v_c/10$.
Using the additive granular drag force model described above, $F = F_z + F_v$, where $F_z \propto f\left(\mu\right) \rho_g g z$ and $F_v \propto \rho_g D^2 v^2$, we would expect the resulting drag force curves at constant velocity to take the form of Fig.~\ref{fig:exp_slow_fast}, \textit{a}, with the addition of a short transient regime~\cite{durian_unified_force}.

As expected, the drag force measured on the intruder is independent of velocity when $v_0<v_c$, as shown in Fig.~\ref{fig:exp_slow_fast}, \textit{b, inset}.
In addition, a short nonlinear regime was observed (Fig.~\ref{fig:exp_slow_fast}, \textit{b, inset}), very similar to that reported in Kang \textit{et al.}~\cite{blumenfeld_nature}.

When increasing the intruder velocity beyond $v_c$, the drag force is expected to deviate systematically from the quasistatic curve in accordance with the predicted velocity dependence (Fig.~\ref{fig:exp_slow_fast}, \textit{a}).
However, when the intruder velocity in experiment is increased above $v_c \sim$ 20 cm/s, surprising behavior results (Fig.~\ref{fig:exp_slow_fast}, \textit{c}).
At shallow penetration depths ($z \lesssim 2d_g$), the nonlinear regime in the quasistatic case (Fig.~\ref{fig:exp_slow_fast}, \textit{b, inset}) has turned into a sharp peak over the same distance that increases with penetration velocity (Fig.~\ref{fig:exp_slow_fast}, \textit{c}, \textit{inset}).
In addition, following the peak in force, tests for which $v_0>v_c$ appear to exhibit drag force deviations above the quasistatic limit for some penetration depth.
At larger depths ($z \gtrsim 5$ cm), all force profiles fall on the trace of the quasistatic experiments, within noise.

In what follows, we will examine these three regimes, namely, the peak in force, the drag force deviations in excess of the quasistatic limit following the peak, and the united, apparently velocity-independent linear regime at greater depths.
We will also situate these regimes in the existing additive drag force framework, identifying ways in which the transition from quasistatic to inertial regimes in granular impact both confirms and questions aspects of this model.
To directly investigate this behavior in terms of system parameters, as well as couple the force felt by the intruder to the grain-scale dynamics, we performed calibrated granular molecular dynamics simulations, described in the following section.

\section{\label{sec:sim_calib}Simulation Calibration}

Though the majority of simulation parameters are taken directly from the experiment, both the friction coefficient $\mu$ of the glass beads and the packing fraction of the bed $\phi_0$ are unknown.
In simulation, both $\mu$ and $\phi_0$ independently affect the resulting strength of the depth dependent drag force, $k_z$, supplying a possible source of ambiguity when attempting a precise calibration.
Indeed, Feng \textit{et al.} find that changing the inter-grain friction coefficient while keeping $\phi_0$ constant changes $k_z$ by more than an order of magnitude~\cite{blumenfeld_soft_matter}.
Conversely, increasing $\phi_0$ while keeping $\mu$ constant also increases $k_z$~\cite{melo_effect_of_pf}.

\begin{figure}[h]
\includegraphics[width=\columnwidth]{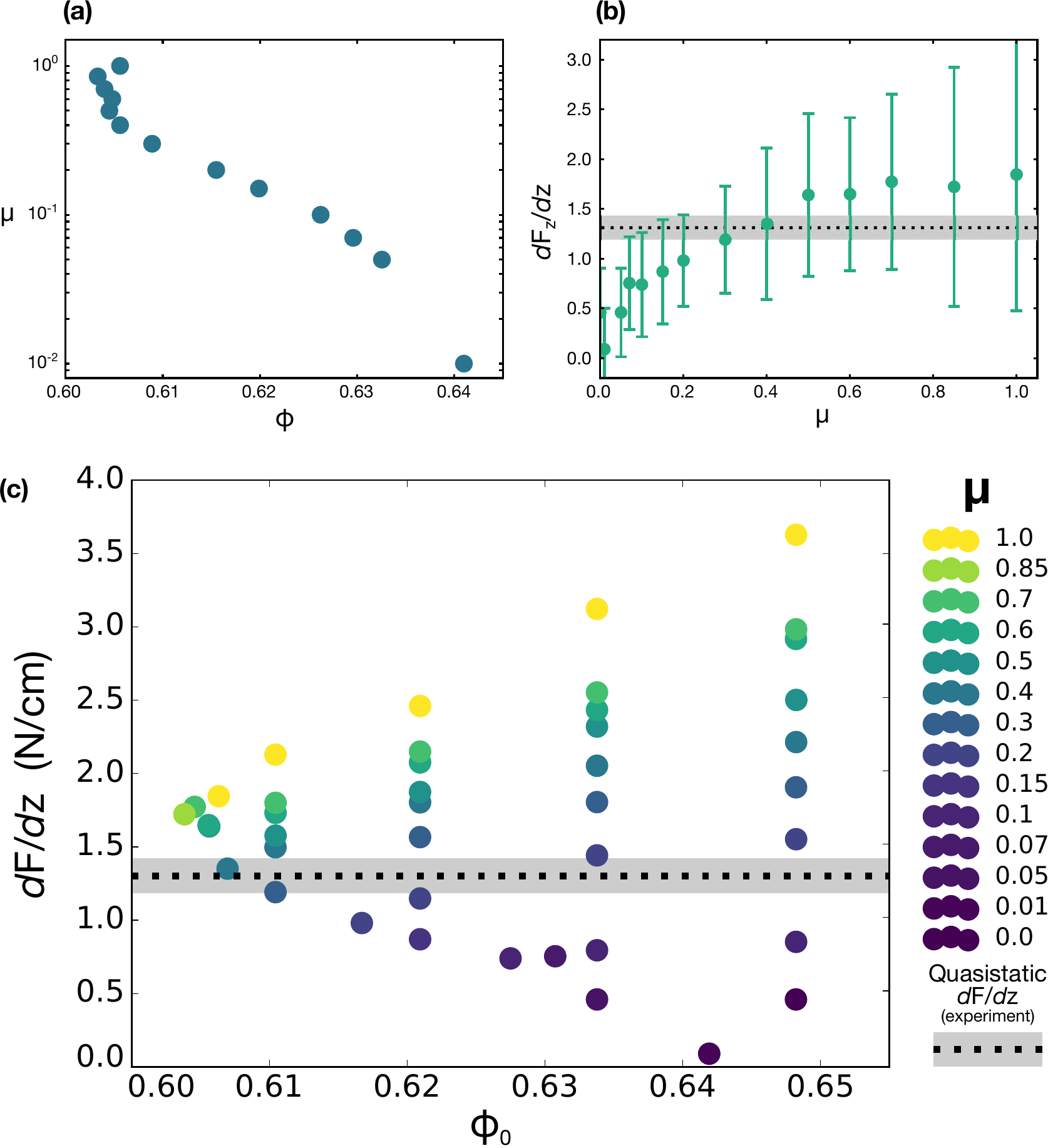}
\caption{\label{fig:sim_pf_calib} \textit{a)} Packing fraction of the bed, $\phi_0$, generated by pouring under gravity while changing $\mu$. \textit{b)} Linear drag force coefficient as a function of the inter-particle friction coefficient $\mu$ for beds prepared via pouring. The simulated intruder's penetration speed is $\sim 2v_c$. \textit{c)} Linear depth dependent drag force for a range of friction coefficient and bed densities, compared with the quasistatic empirical result. The simulated intruder's penetration speed is $\sim 2v_c$.}
\end{figure}

As has been previously observed, when using the simulated `pouring' bed preparation method as described above, $\phi_0$ is a function of $\mu$~\cite{silbert_mu} (Fig.~\ref{fig:sim_pf_calib}, \textit{a}).
Given a stable bed generated at low $\mu$, such that it has a high $\phi_0$, one may then increase $\mu$ in the poured bed without fear of destabilizing the aggregate, whereas lowering $\mu$ for a given configuration could produce rearrangements.
It is thus possible to generate an initial bed configuration with a given $\mu$ at a stable $\phi_0$, and then increase $\mu$ in the static packing to yield a suite of different values of $k_z$.
With this procedure (Fig.~\ref{fig:sim_pf_calib}, \textit{c}), it appears that there is a continuous range of $\left(\mu,\phi_0 \right)$ pairings that would produce a value of $k_z$ consistent with the quasistatic limit in the experiment.

We have here decided to avoid this technique, as it may produce beds whose grain-scale properties, frictional properties, and force networks do not correspond to an experimentally poured system~\cite{silbert_mu}.
In addition, we avoid the ambiguity inherent in choosing a specific pairing of $\left(\mu,\phi_0 \right)$ out of the many that produce the same value of $k_z$, but which may affect auxiliary behavior, such as the peak or other transient effects.
Instead, we allow the interaction between $\mu$ and the preparation conditions --- here, pouring under gravity --- to determine the initial packing fraction of the bed, as in Fig.~\ref{fig:sim_pf_calib}, \textit{a}.
This procedure essentially removes $\phi_0$ as an independent variable, while also ensuring that the force networks within the bed were produced in the same conditions under which they will be tested.
Thus, in what follows, $\mu$ is the principal parameter that controls both $\phi_0$, via the bed preparation conditions, as well as the strength of the depth dependence, $k_z$, which depends on both $\phi_0$ and $\mu$~\cite{blumenfeld_soft_matter}.

A friction coefficient $\mu=0.3$ has been used previously for simulations of glass sphere aggregates~\cite{DEM_sims_mu,tanaka_plate_drag,makse_mu_03}.
As can be seen in Fig.~\ref{fig:sim_pf_calib}, \textit{b}, $\mu = 0.3$ yields agreement between the simulation and experimental quasistatic result, and the corresponding packing fraction of the simulated bed, $\sim 0.61$, is also realistic.
In the following discussion, $\mu$ is fixed at 0.3 in simulation.

\begin{figure}[h]
\includegraphics[width=\columnwidth]{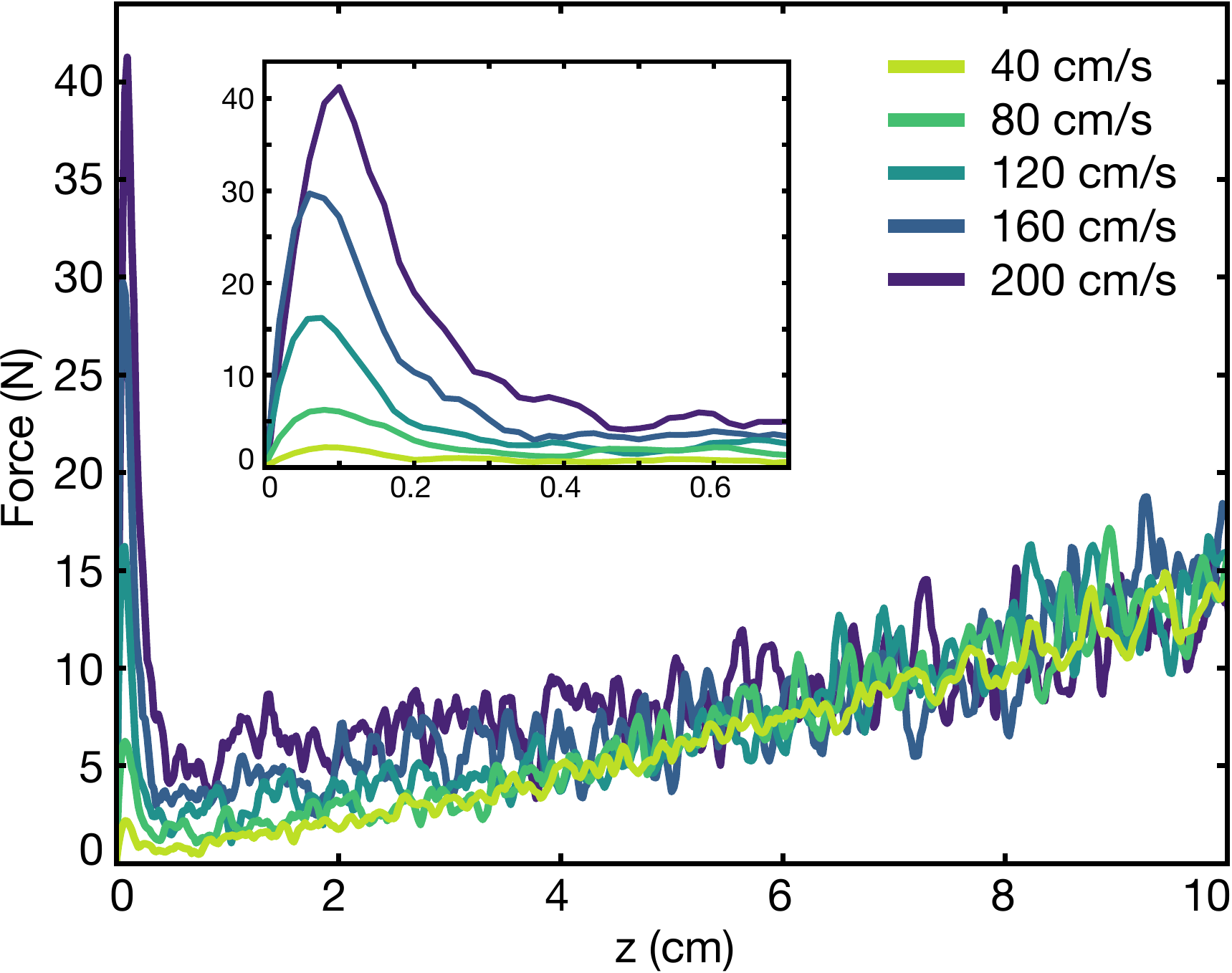}
\caption{\label{fig:sim_only_full} Drag force as a function of penetration depth in simulation, for selected velocities. \textit{Inset:} Detail of the transient force peak.}
\end{figure}

We performed penetration simulations with the parameters described above for impact velocities ranging from 40 to 240 cm/s, with the aim of matching all three effects observed in the experiment: the peak, the velocity-dependent transient, and the linear regime.
A selection of these results is shown in Fig.~\ref{fig:sim_only_full}.

Qualitatively, the drag force measured in simulation is very similar to the experimental results in Fig.~\ref{fig:exp_slow_fast}, \textit{c}.
As $v_0$ increases, the transient force peak grows in height and the post-peak drag force increasingly decouples from the force curve representing the quasistatic limit, in this case, 40 cm/s ($v_c \sim 20$ cm/s), before all curves unite at $z \gtrsim 6$ cm.

\begin{figure}[h]
\includegraphics[width=\columnwidth]{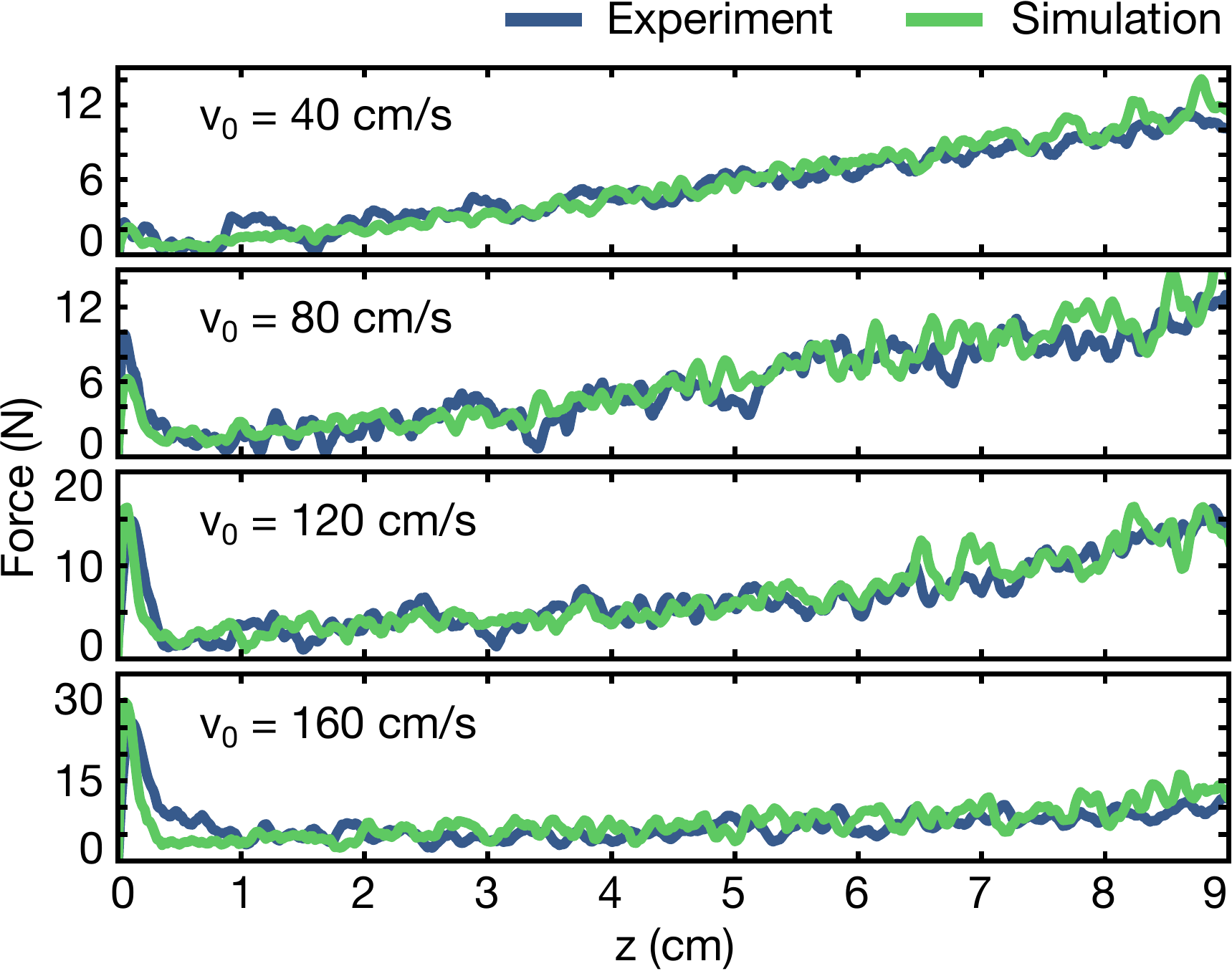}
\caption{\label{fig:sim_exp_comp_full} Simulated and experimental drag force as a function of penetration depth for selected velocities.}
\end{figure}

\begin{figure}[h]
\includegraphics[width=\columnwidth]{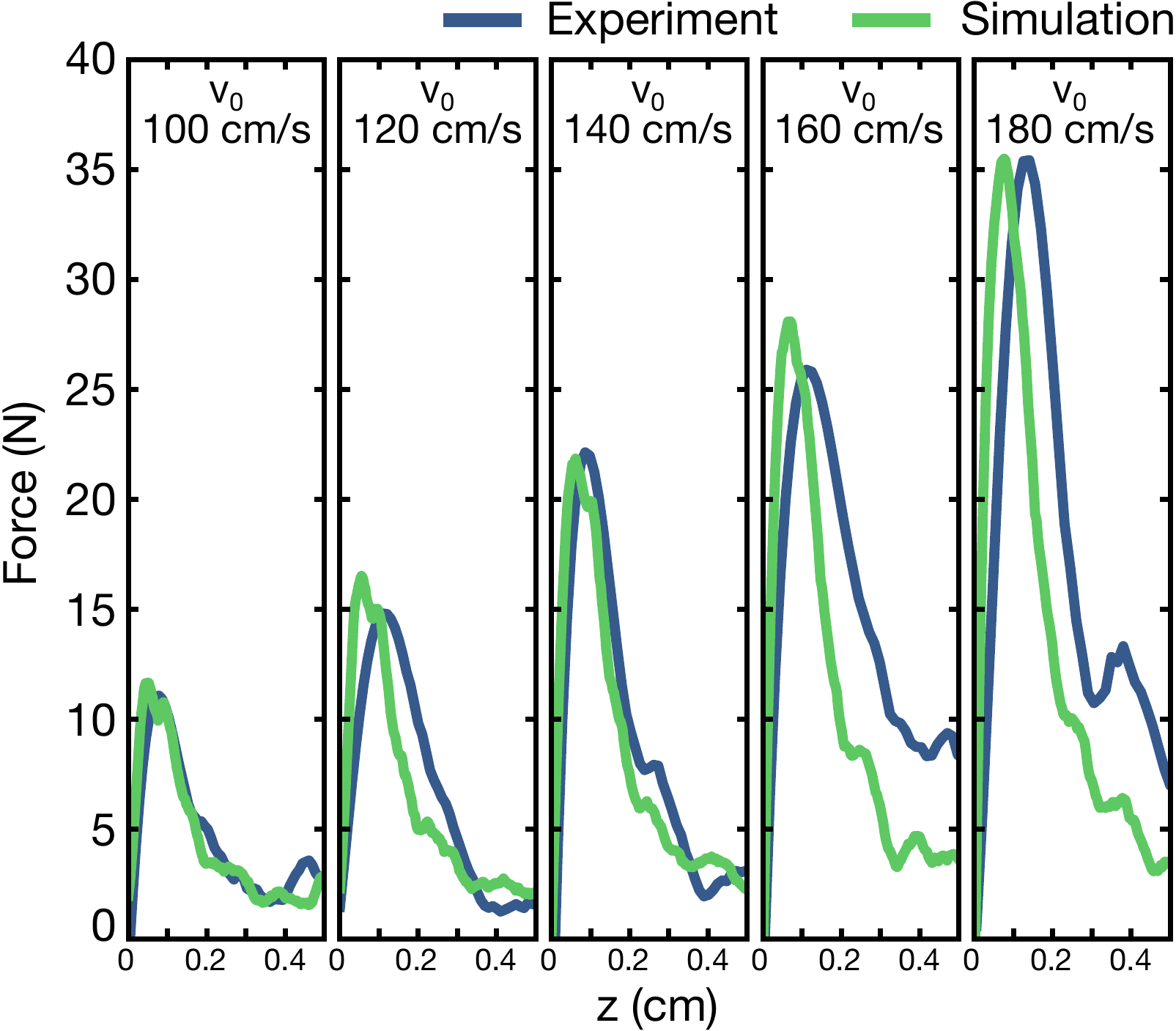}
\caption{\label{fig:sim_exp_comp_peak} Comparison of the simulated and experimentally observed transient force peaks as a function of penetration depth for selected velocities.}
\end{figure}

\begin{figure}[h]
\includegraphics[width=\columnwidth]{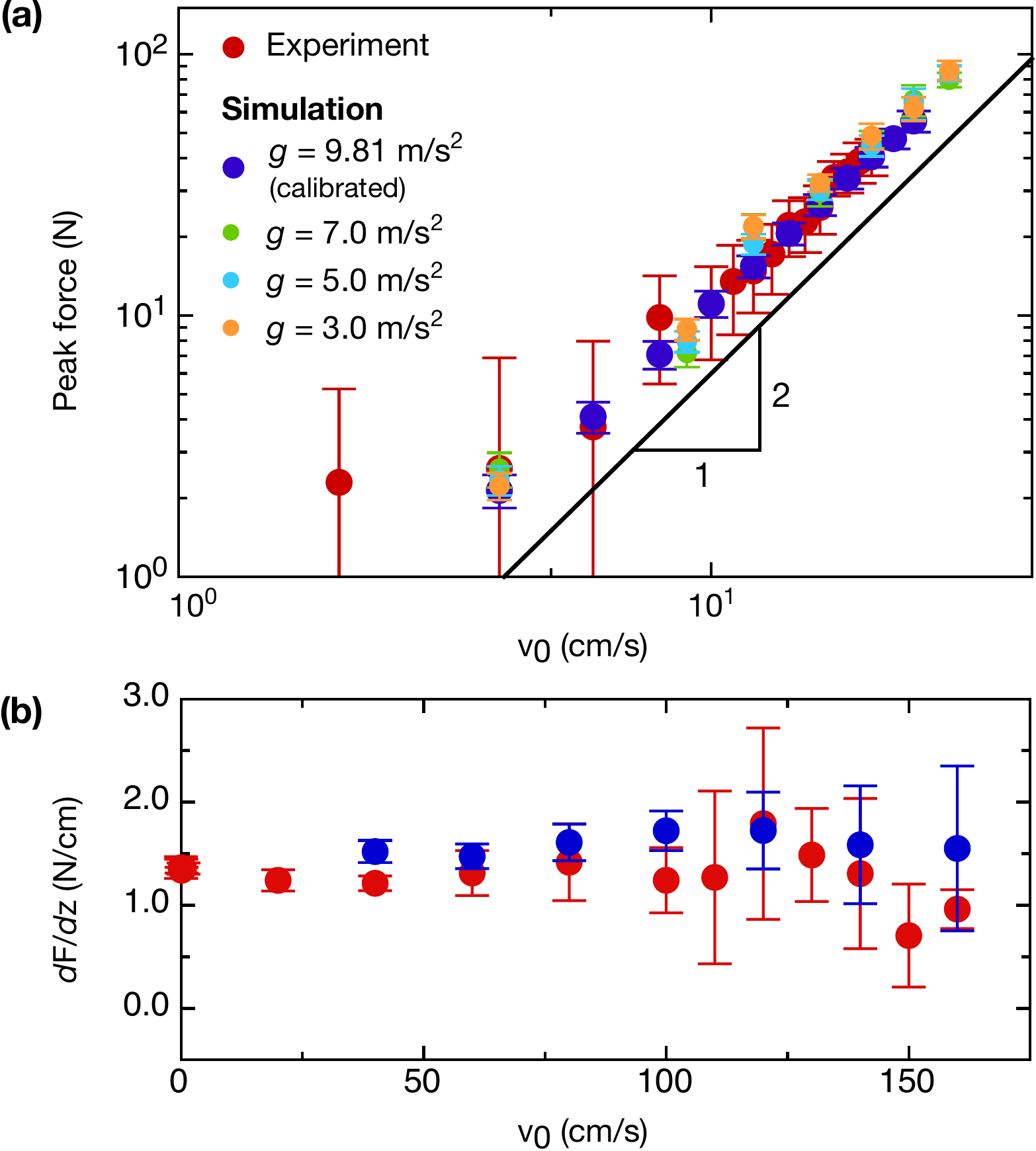}
\caption{\label{fig:sim_exp_peak_and_slope} \textit{a)} Peak force, from the experiment, calibrated simulation, and simulations with different values of $g$, on a log scale as a function of intruder velocity with a $v_0^2$ slope plotted as a guide. \textit{b)} Linear fits of the slope at $z \geq 3$ cm, plotted for experiment and simulation.}
\end{figure}

The agreement between the simulations and experiment is demonstrated by the direct comparisons in Fig.~\ref{fig:sim_exp_comp_full}.
The fluctuations seen in Fig.~\ref{fig:exp_slow_fast}, \textit{c} are also found in the simulation results, and both have an approximate wavelength of $d_g = 2$ mm.
These fluctuations are likely due to grain-level rearrangements during impact, and may decrease in relative magnitude if the intruder cross sectional area increased.

A comparison between the transient force peak in simulation and experiment is shown in Fig.~\ref{fig:sim_exp_comp_peak}.
Though the peak is extremely short --- it persists for only a few milliseconds during impact --- it is well captured by the simulations in profile and magnitude.
A quantitative comparison of transient peak height and the strength of the depth dependence is found in Fig.~\ref{fig:sim_exp_peak_and_slope}.
There is excellent agreement between the peak heights found in the experiment and simulation (Fig.~\ref{fig:sim_exp_peak_and_slope}, \textit{a}), and both are consistent with a $v_0^2$ scaling.
In addition, despite large fluctuations, the slope of the depth dependence $k_z$ is independent of velocity and shows agreement between experiment and simulation (Fig.~\ref{fig:sim_exp_peak_and_slope}, \textit{b}).

Because the only fitting parameter between the simulation and experiment was $\mu$, which then determined $\phi_0$ via the preparation conditions, the strong agreement indicates that we are capturing the salient physical parameters.
Armed with this agreement in both the long and the short timescales, both the velocity dependent and quasistatic characteristics, we will examine the grain scale dynamics accessible in simulation and correlate these with the drag force data.
In this way, we aim to shed light on the mechanisms underlying this diverse behavior and how it relates to existing models.

\section{\label{sec:depth_dependence}Depth Dependence}

\begin{figure}[h]
\includegraphics[width=\columnwidth]{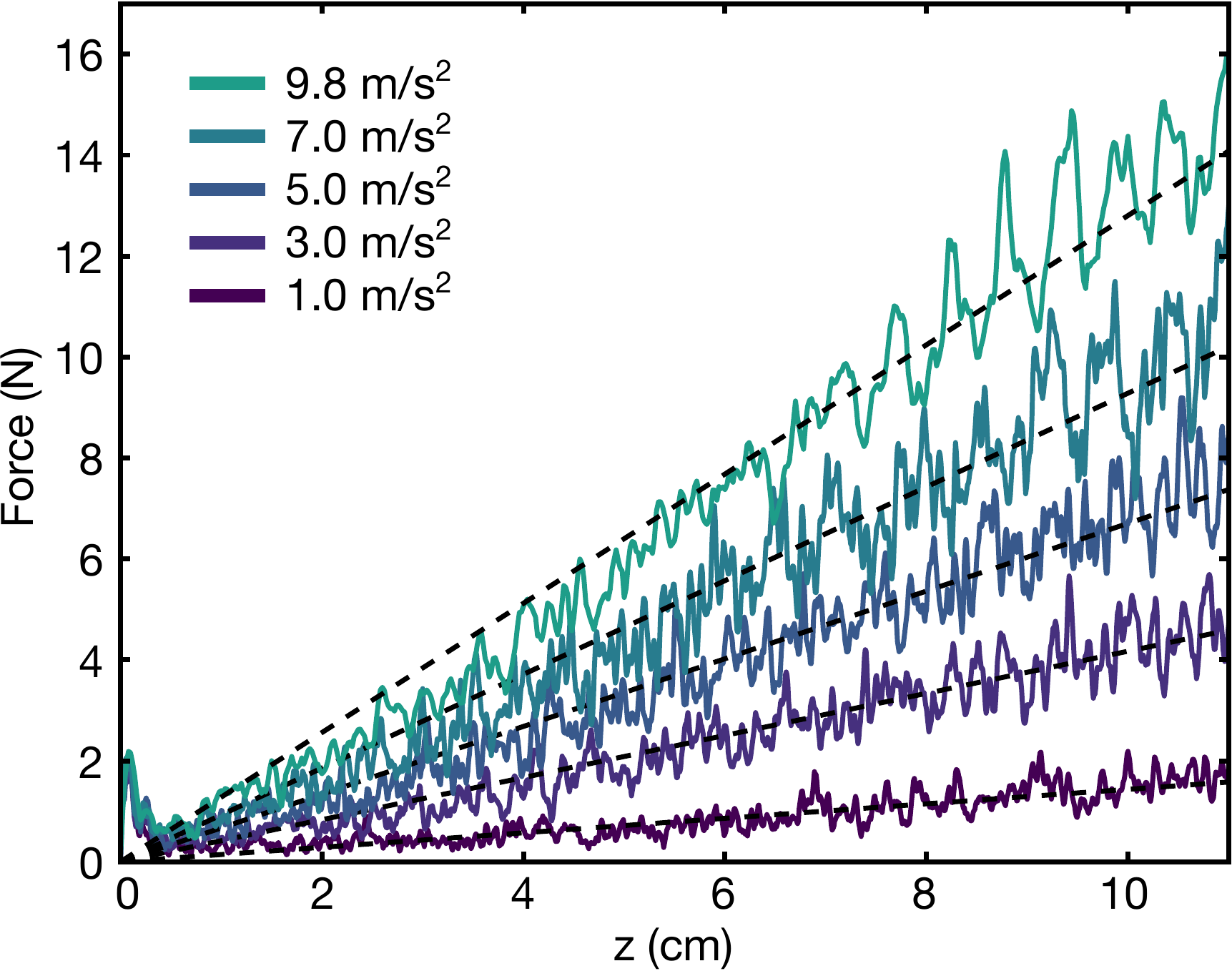}
\caption{\label{fig:slopes_40cms_full} Simulated force as a function of depth at different values of gravity, for $v_0=40$ cm/s and $D=2$ cm. The bed configuration was kept constant when changing $g$. Linear fits are plotted in black.}
\end{figure}

In Fig.~\ref{fig:slopes_40cms_full}, we calculate the depth dependent force extracted via linear fit in the relatively low speed penetration simulations ($v_0 = 40$ cm/s) and explicitly probe the dependence of $F_z$ on gravity.
We follow the functional form proposed by Brzinski III \textit{et al.}, as described in Section~\ref{sec:intro}: $F_z = \alpha \mu \rho_g g D^2 z$, where $\alpha$ was found to be approximately $\sim 20-30$~\cite{durian_depth_dependent,durian_unified_force}.

The fitted values of $\alpha$ are $35 \pm 5$, and do not vary strongly with gravity.
These values are nearly within error of those reported by Brzinski III \textit{et al.}, without accounting for disparities in packing fraction and friction.
Strength is further lent to this agreement due to the difference in experimental procedure: where Brzinski III \textit{et al.} measured the depth dependent force in static conditions, we extract our from penetration at speeds slightly higher than the quasistatic limit, as evidenced by the peak at low $z$ in Fig.~\ref{fig:slopes_40cms_full}~\cite{durian_depth_dependent}.

\section{\label{sec:vel_dependence}Velocity Dependence}

\begin{figure}[h]
\includegraphics[width=\columnwidth]{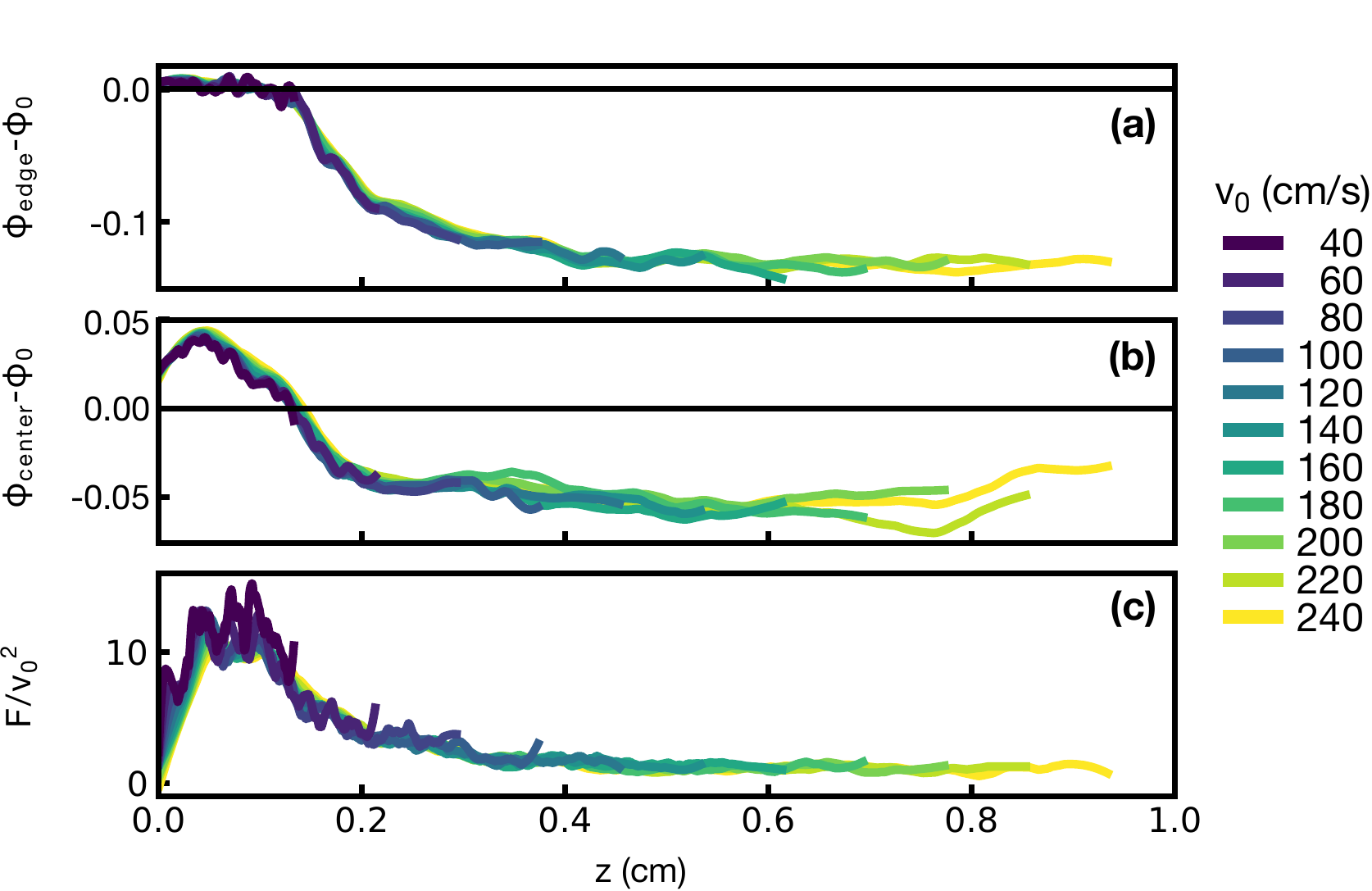}
\caption{\label{fig:peak_pf} \textit{a)} Change in packing fraction at the edge of the intruder, as a function of penetration depth. \textit{b)} Change in packing fraction just below the center of the rod surface. \textit{c)} Peak in force, normalized by the square of the impact velocity.}
\end{figure}

The force peak appears to represent the process of granular yielding.
In Fig.~\ref{fig:peak_pf}, \textit{c}, the force peak, when normalized by the square of the impact velocity, as indicated in Fig.~\ref{fig:sim_exp_peak_and_slope}, collapses as a function of depth.
Gravish \textit{et al.} also observed some velocity dependence on yielding in granular drag situations ~\cite{goldman_force_flow_pre}.
The packing fraction in the center of the rod surface initially increases (Fig.~\ref{fig:peak_pf}, \textit{b}), reaching a maximum roughly at the same $z$ value as the force.
As the force begins to decay from its maximum value, $\phi_{\mathrm{center}}$ begins to decrease as well, indicating the beginning of flow.
As $\phi_{\mathrm{center}}$ descends below the original bed density ($\phi_{\mathrm{center}} -\phi_0 = 0$), the force begins to level off and the packing fraction at the edge of the rod, $\phi_{\mathrm{edge}}$, begins to decrease rapidly (Fig.~\ref{fig:peak_pf}, \textit{a}).
The invariant position of the force peak for all impact velocities tested here, as well as the transition from grain compression to dilation and flow, is consistent with the force characteristics of yielding observed in~\cite{goldman_force_flow_pre}.

\begin{figure}[h]
\includegraphics[width=\columnwidth]{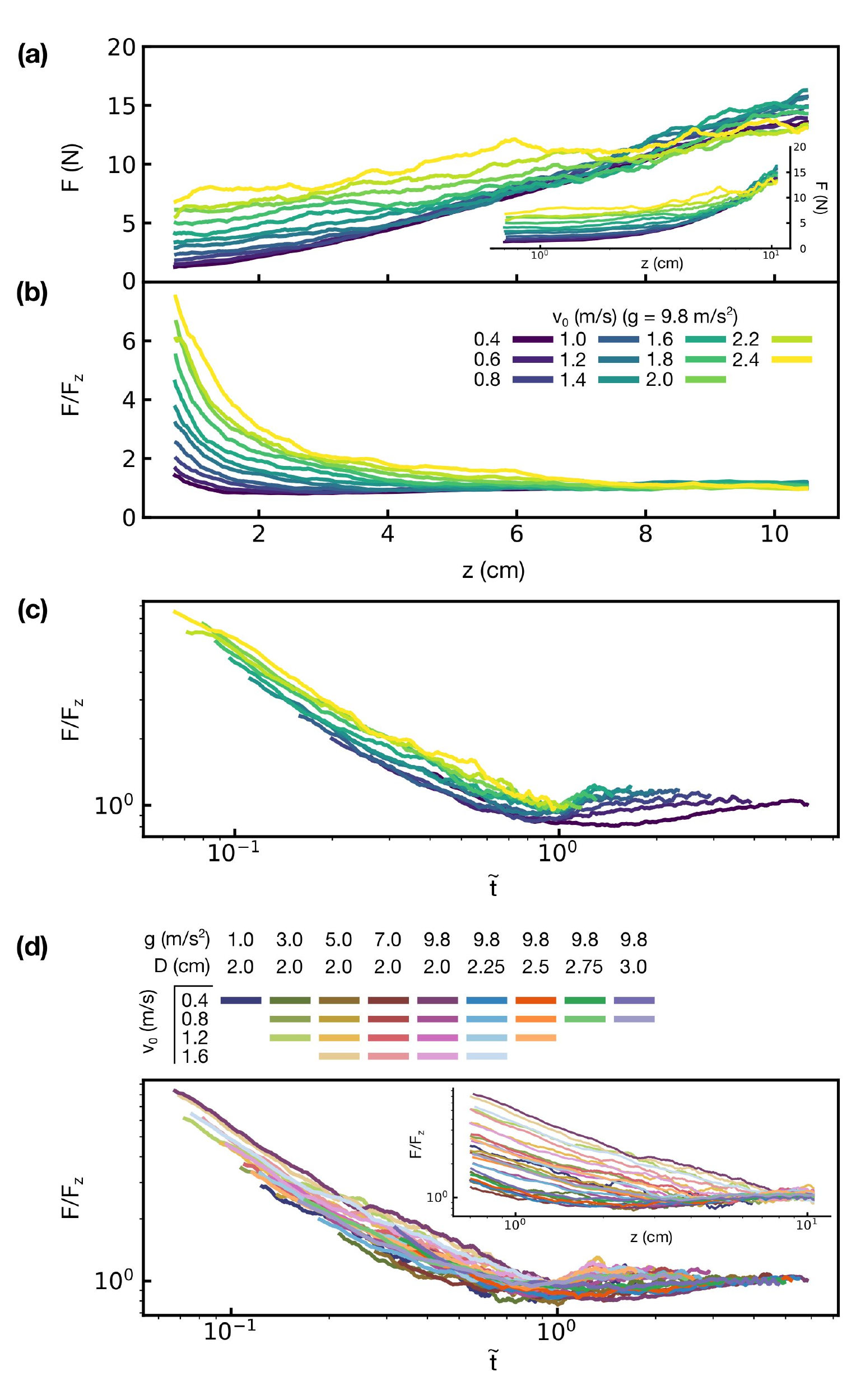}
\caption{\label{fig:force_scaling} \textit{a)} Force on the rod as a function of depth, on a linear and (\textit{inset}) log scale. \textit{b)} Force on the rod divided by the fitted depth dependent force $F_z$. In the linear regime, $F/F_z=1$. \textit{c)} $F/F_z$ as a function of the rescaled time, $\widetilde{t} = \left(z/v_0\right)\sqrt{g/D}$. \textit{d)} Normalized force as a function of $\widetilde{t}$ for different values of $v_0$, $g$ and $D$. The parameter $\widetilde{t}$ collapses the transition from velocity dependent to velocity independent drag force behavior.}
\end{figure}

In Fig.~\ref{fig:exp_slow_fast}, \textit{c} and Fig.~\ref{fig:sim_exp_comp_full}, there are some transient effects at low $z$ following yielding.
In Fig.~\ref{fig:force_scaling}, we plot again the simulated drag force as a function of rod position $z$ for the full range of impact speeds tested, excluding the initial peak.
The strong dependence of the strength of the transient on $v_0$ suggests that this behavior could be a manifestation of the force $F_v$ described above.
Based on projectile experiments, Katsuragi and Durian find that $F_v$ is given by $0.8 \rho_g D^2 v^2$, where, in the case of constant velocity, $v = v_0$~\cite{durian_unified_force}.

In order to isolate the velocity dependent transient, we divide the drag force $F$ by $F_z = \alpha \mu \rho_g g D^2 z$, using the values of $\alpha$ fitted earlier (Fig.~\ref{fig:force_scaling}, \textit{b})
In our system, however, the transient eventually intersects with the quasistatic limit (Fig.~\ref{fig:force_scaling}, \textit{a}, \textit{inset} and \textit{b}), and the point at which each force curve, $F$ connects with $F_z$ appears to depend on $v_0$.
Because this intersection point depends on velocity, we define a dimensionless time variable:

\begin{equation}
    \widetilde{t} = \frac{z}{v_0} \sqrt{\frac{g}{D}}
\end{equation}

The normalized drag force is plotted as a function of $\widetilde{t}$ in Fig.~\ref{fig:force_scaling}, \textit{c}.
At early times, the drag force is in excess of $F_z$ by an amount that is continuously decreasing with time, until, at $\widetilde{t}=1$, $F$ abruptly joins $F_z$.
To directly test the quality of the nondimensionalization of time, in Fig.~\ref{fig:force_scaling}, \textit{d}, we plot the normalized force for five different values of $g$ and four values of $D$.

Though there is scatter, there does not appear to be a systematic trend that would conflict with this normalization, and the transition from $F$ to $F_z$ at $\widetilde{t}=1$ is well captured.
This robust transition suggests that, as $v_0$ exceeds $v_c$, there is a timescale in the flow which is mediated by gravity and that qualitatively --- and temporarily --- changes the nature of the force experienced by the rod.

In what follows, we investigate the grain scale effects occurring around the intruder in simulation in an effort to provide a physical rationale for the timescale $\widetilde{t}$, focusing in particular on the signatures of the transition from velocity dependent to velocity independent drag force.
We do not hereafter include the initial yielding peak in the intruder force trace, given that it occurs at penetration depths less than $0.7 R$, where $R =D/2$ is the radius of the intruder (see Fig.~\ref{fig:peak_pf}, \textit{c}), and that it presages the development of the flow fields examined below.

\section{\label{sec:shear_zone}Inertial Number}

A gravity dependent timescale arbitrating the transition from velocity dependent to velocity independent forces implies that inertial effects, and therefore flow characteristics, may be important only at early times.
Because, on average, there is no circular flow about the $\hat{z}$ axis, we can average dynamic grain quantities in the azimuthal direction and plot them as a function of $\left(r,z\right)$ with respect to the axis of the rod.

\begin{figure}[h]
\includegraphics[width=\columnwidth]{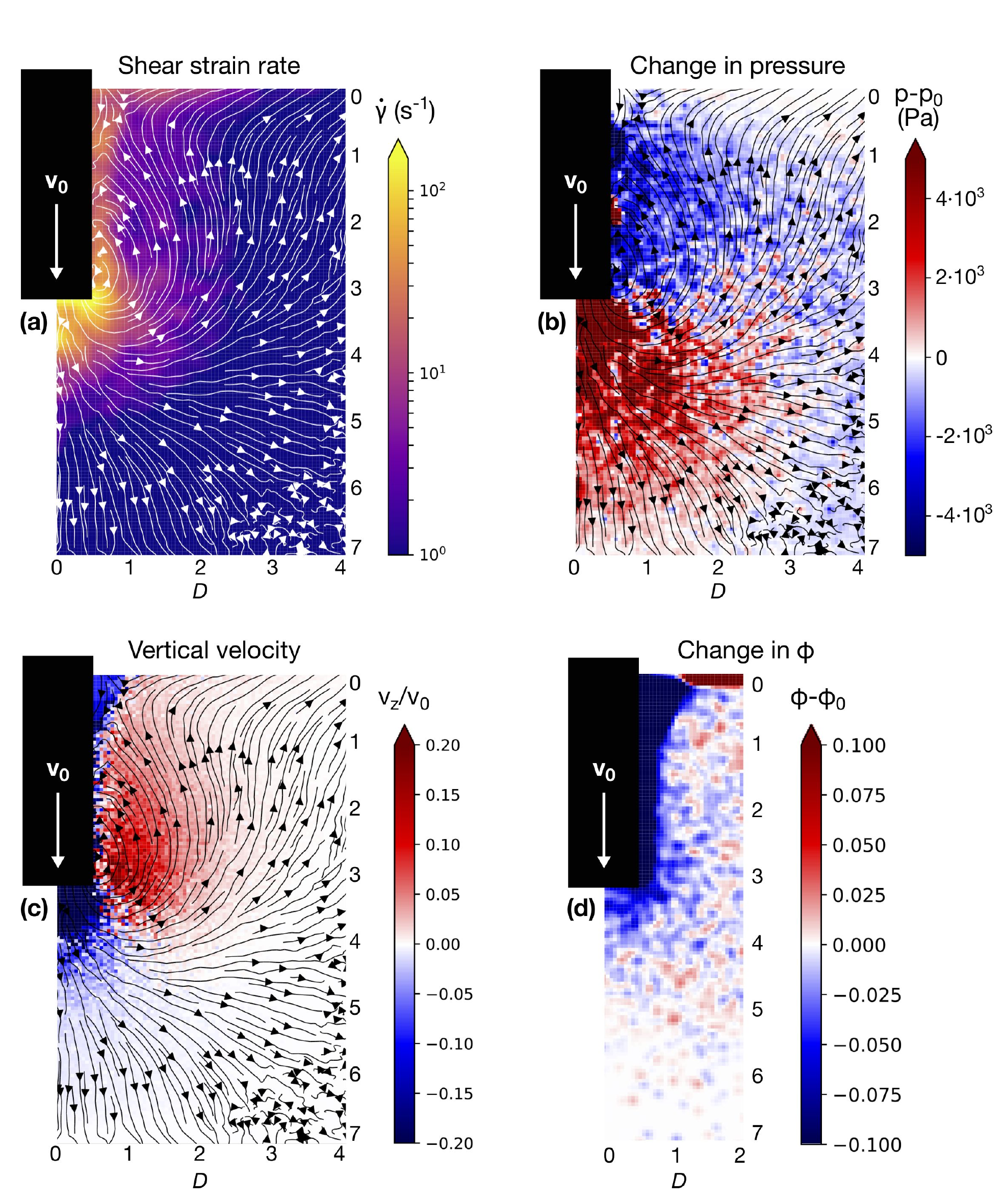}
\caption{\label{fig:heatmaps} Stream plots and flow fields for $v_0 = 140$ cm/s at scaled time $\widetilde{t}=1.0$. \textit{a)} Shear strain rate $\dot\gamma$ is focused around the edge of the rod. \textit{b)} Change in grain scale pressure field, $p-p_0$, during impact. \textit{c)} Vertical component of the grain velocity, $v_z$. \textit{d)} The change in packing fraction of the grains surrounding the intruder, $\phi-\phi_0$. A negative value indicates dilation.}
\end{figure}

Using the local, grain scale velocity field we calculate the streamlines of the flow, as well as the maximal shear strain rate, $\dot{\gamma}$, given by:

\begin{equation}
    \dot{\gamma} = \sqrt{\frac{1}{2}\left(\frac{\partial v_r}{\partial r}-\frac{\partial v_z}{\partial z}\right)^2 + \frac{1}{2}\left(\frac{\partial v_r}{\partial z}+\frac{\partial v_z}{\partial r}\right)^2}
\end{equation}

where $v_r$ and $v_z$ are the radial and vertical components of the grain velocity~\cite{goldman_robo}.
We also calculate the local volume fraction, $\phi$, with a Voronoi transform.
Fig.~\ref{fig:heatmaps} gives an overview of the general flow characteristics at $\widetilde{t} = 1$.

Some features of the flow stand out.
The shear strain rate is focused at the edge of the rod and along the rod shaft, but also extends into a region below the intruder (Fig.~\ref{fig:heatmaps},\textit{a}).
It is likely that, if the ratio of grain size to rod diameter was much smaller, there would be a well-defined shear band that extended from the edge of the rod, forming a triangle, as in~\cite{tanaka_plate_drag,melo_dynamics_shear,goldman_force_flow_pre}.

Areas of high pressure are concentrated under the rod surface and radiate outwards at an angle that is likely related to the internal friction angle (Fig.~\ref{fig:heatmaps}, \textit{b}), and the vertical component of the grain velocity exhibits a similar angle demarcating the circulation of the flow (Fig.~\ref{fig:heatmaps}, \textit{c}).
Above the rod surface, however, the pressure is significantly lowered, possibly indicating a degree of fluidization.

The change in local packing density behaves in a similar way to the shear strain rate and grain velocity.
Because $\phi_0 = 0.61$ is above $\phi_c$, which is generally 0.59-0.60 for glass spheres, the grains must dilate before they can flow~\cite{goldman_gran_imp_2010, goldman_force_flow_pre}.
At this impact speed, the packing fraction directly under the surface of the rod decreases by 5-10\% (Fig.~\ref{fig:heatmaps}, \textit{d}).

\begin{figure}[h]
\includegraphics[width=\columnwidth]{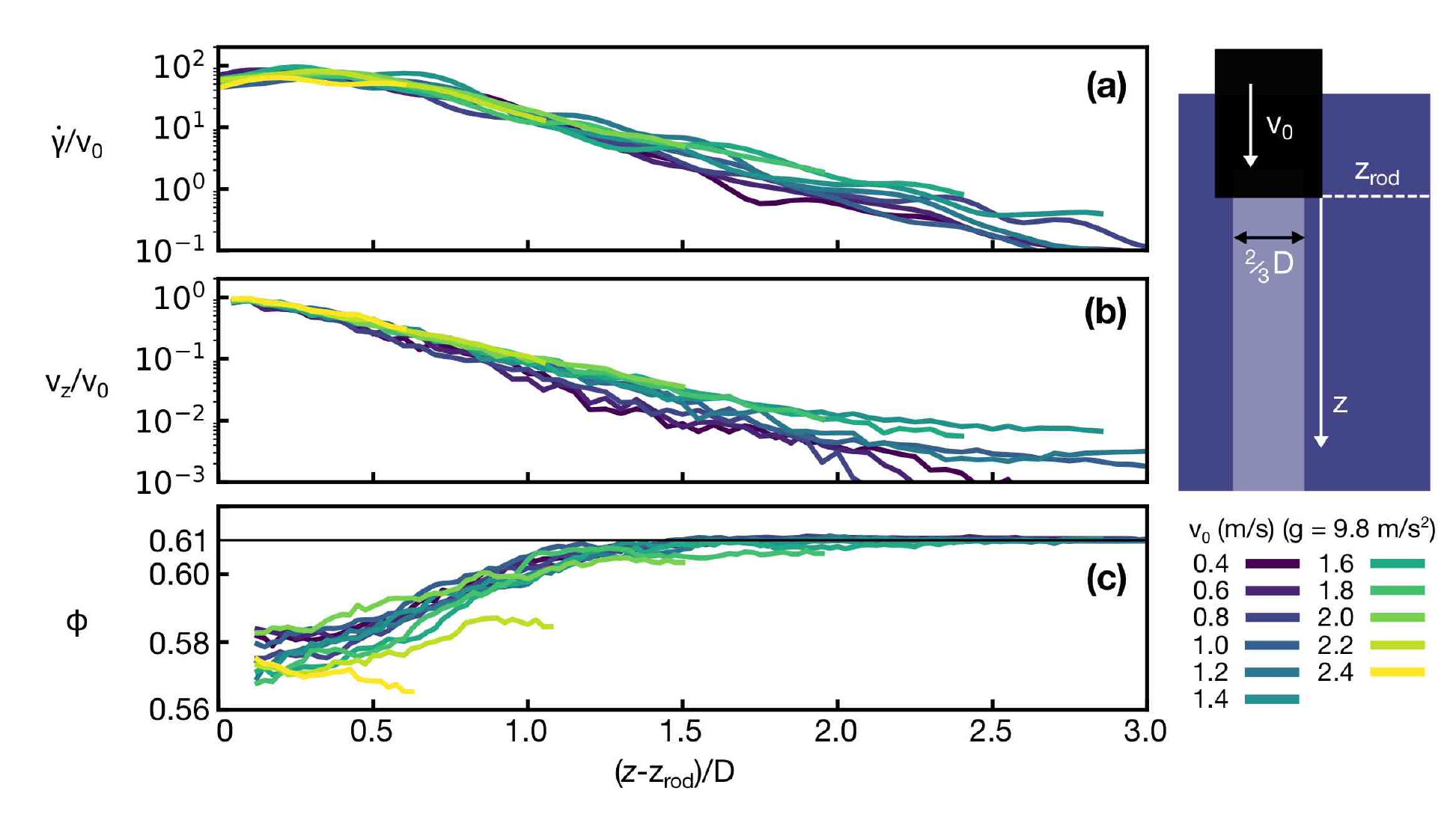}
\caption{\label{fig:gam_vz_pf_prof} Flow profiles averaged within an cylinder of radius 2/3 $D$ aligned with the intruder axis. Each profile is taken at $\widetilde{t} = 1.0$. \textit{a)} Shear strain rate $\dot \gamma$ as a function of distance from the rod surface. \textit{b)} Vertical grain velocity profile. \textit{c)} Local packing density under the rod surface.}
\end{figure}

To investigate the flow characteristics below the rod surface more quantitatively, in Fig.~\ref{fig:gam_vz_pf_prof} we plot vertical profiles of $\dot{\gamma}$, $v_z$ and $\phi$, averaged in an cylinder of diameter $2/3 D$ (Fig.~\ref{fig:gam_vz_pf_prof}), at the same $\widetilde{t}$.
This area, highlighted in Fig.~\ref{fig:gam_vz_pf_prof}, \textit{right}, was chosen to avoid the areas of high shear around the edge of the intruder.

The $\dot{\gamma}$ and $v_z$ profiles show clear exponential behavior that extends several $D$ away from the intruder surface, consistent with the experiments of Seguin \textit{et al.}~\cite{gondret_exp_vel} (Fig.~\ref{fig:gam_vz_pf_prof}, \textit{a} and \textit{b}).
Additionally, the local packing density in this inner cylinder is lowered by about 3\% near the intruder surface, independent of $v_0$ (Fig.~\ref{fig:heatmaps}, \textit{c}).
The dilation region ends uniformly at $\sim D$ away from the rod, very close to the mean lengthscale $\lambda_{vz} = 0.98 D \pm 0.15$ extracted via exponential fit from the $v_z$ profile, which is independent of $v_0$.
Under these conditions, there is no evidence of an underlying jammed structure or stagnant zone such as those suggested by Aguilar and Goldman~\cite{goldman_robo} or Kang \textit{et al.}~\cite{blumenfeld_nature}.

The flow characteristics in the area directly below the rod are regular and directly proportional to $v_0$, and do not exhibit an obvious signature of a transition between inertial and quasistatic behaviors.
However, the region of most violent shear --- and the highest degree of inertial flow --- is contained in the flow around the bottom edge of the rod (as can be seen in Fig.~\ref{fig:heatmaps}, \textit{a}).
To investigate the possibility of a force transition signature in the area of peak flow, the value of $\dot{\gamma}$, $\phi$ and the grain pressure at $r=R$ are averaged and recorded as a function of intruder depth.
We use the grain pressure and shear strain rate to calculate the inertial number $I$, given by:

\begin{equation}\label{eq:inertial}
    I = \frac{\dot{\gamma} d_g}{\sqrt{P/\rho_g}}
\end{equation}

where $P$ is the pressure and $d_g$ is the diameter of a grain~\cite{pouliquen_flows_of_dense,singh_inertial}.
The inertial number is used to compare the timescale associated with deformation of the material, given by $\dot{\gamma}$, with the rearrangement time under a pressure $P$~\cite{pouliquen_flows_of_dense}.
Small values of $I \leq 0.01$ correspond to quasistatic flow, whereas $I \geq 0.1$ delimits the collisional regime and the presence of inertial effects~\cite{pouliquen_flows_of_dense,koval_inertial_num}.

\begin{figure}[h]
\includegraphics[width=\columnwidth]{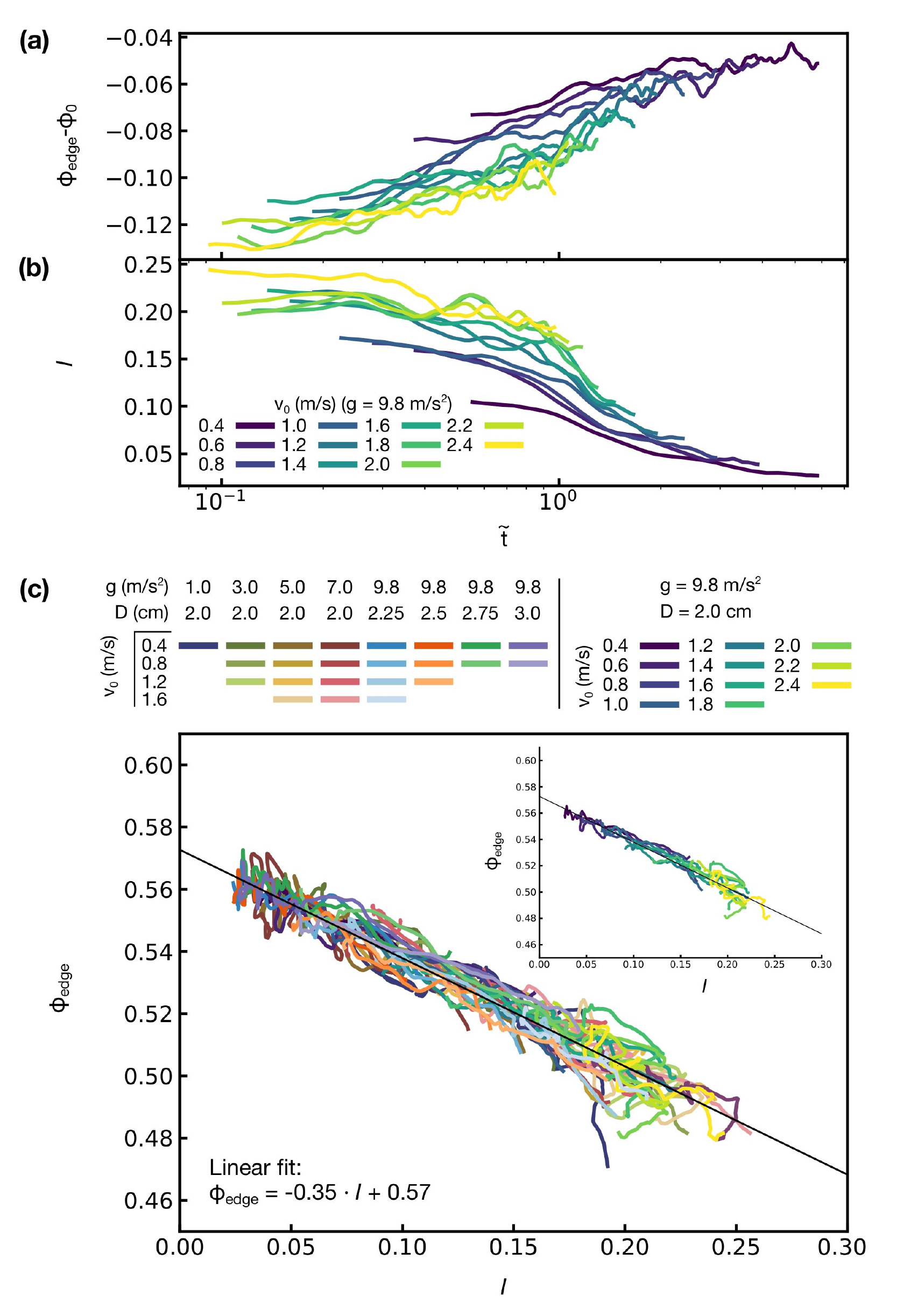}
\caption{\label{fig:inertial_num} \textit{a)} Change in packing fraction in the region near the edge of the rod as a function of $\widetilde{t}$ in a co-moving frame. \textit{b)} Inertial number of the flow in the same region. \textit{c)} $I$ versus $\phi_{\mathrm{edge}}$ for a range of values of $g$ and $D$, plotted with a linear fit. \textit{Inset:} $\phi_{\mathrm{edge}}$ plotted against $I$ for only the simulations calibrated with the experimental data.}
\end{figure}

In Fig.~\ref{fig:inertial_num}, \textit{b}, we plot the inertial number at the edge of the rod as a function of $\widetilde{t}$, and in \textit{a} is the change in packing fraction at the edge of the rod, $\phi_{\mathrm{edge}}-\phi_0$.
There is scatter, but some general observations can be drawn.
Our general results for the inertial number are consistent with $I$ as measured empirically by Clark \textit{et al.}~\cite{clark_steady_flow_dynamics}.
Clark \textit{et al.} calculated the inertial number for a circular intruder impacting a system of disks, finding that $I \sim 0.09$ in the area directly below the intruder for impact speeds below 6 m/s.

When $\widetilde{t}=1$, the rates of change of both $I$ and $\phi_{\mathrm{edge}}-\phi_0$ increase markedly.
In addition, when $\widetilde{t}<1$, $\phi_{\mathrm{edge}}$ is increasing (and $I$ decreasing), albeit at a more gradual rate.
This may be indicative of the hybrid nature of the flow as a function of depth: the gradual trends in $I$ and $\phi_{\mathrm{edge}}$, especially for higher $v_0$, reflect that the flow is operating in increasing ambient pressure conditions, so that there is some bleeding between the `inertial,' high $v_0$ regime, and the later quasistatic behavior.
Nonetheless, it seems that we are able to draw the conclusion, from Fig.~\ref{fig:inertial_num} \textit{a} and \textit{b}, that the gradual change of ambient static granular conditions is not sufficient to explain the bend in $I$ and $\phi_{\mathrm{edge}}$ at $\widetilde{t}=1$.

Granular rheology experiments and simulations under simple flow configurations have shown that $I$ and $\phi$ are related in the following way:

\begin{equation}
    \phi \approx \phi_{\mathrm{ini}} - mI
\end{equation}

where $m$ is a constant and $\phi_{\mathrm{ini}}$ is the initial packing fraction, usually given as 0.2 and 0.6 for glass spheres in three dimensions, respectively~\cite{pouliquen_flows_of_dense,chevoir_rheophysics_dense,andrade_inertial_flows}.
To compare, we plot $\phi$ and $I$ in Fig.~\ref{fig:inertial_num}, \textit{c}.
The best fit line is given by $\phi = 0.57-0.35I$, plotted in black \textit{c}, and is remarkably close to the prediction from granular rheology --- especially for such irregular and dynamic flow conditions.
Interestingly, $\phi_{\mathrm{ini}}$ is much smaller than both the initial packing fraction of the system, $0.61$, and also the predicted dilation threshold, $\phi_c \sim 0.60$~\cite{goldman_force_flow_pre}.
However, Kobayakawa \textit{et al.} have observed in their simulations of horizontal drag that the packing density within a shear band, $\phi_s$ can be lower than $\phi_c$, and they find that $\phi_s = 0.58$ to $0.59$ in their simulations~\cite{tanaka_plate_drag}.
The average packing fraction in the center of the rod, away from the edge, as plotted in Fig.~\ref{fig:gam_vz_pf_prof}, \textit{c}, indicates that $\phi_s$ for this system is near 0.58, agreeing well with Kobayakawa \textit{et al.}.

The linearity of Fig.~\ref{fig:inertial_num}, \textit{c}, indicates that the flow around the rod surface is regular and well captured by existing models; in other words, there is not a discontinuity in flow at the transition between velocity and depth dependent regimes.
Though there are signatures of the transition in $I$ and $\phi_{\mathrm{edge}}$ at $\widetilde{t}=1$, it would seem that this is due to additional actors in the system that influence the flow but do not disturb it.
To explain the force regime transition, then, we must look elsewhere.

\section{\label{sec:temporal_flow}Temporal Flows}

\begin{figure}[h]
\includegraphics[width=7cm]{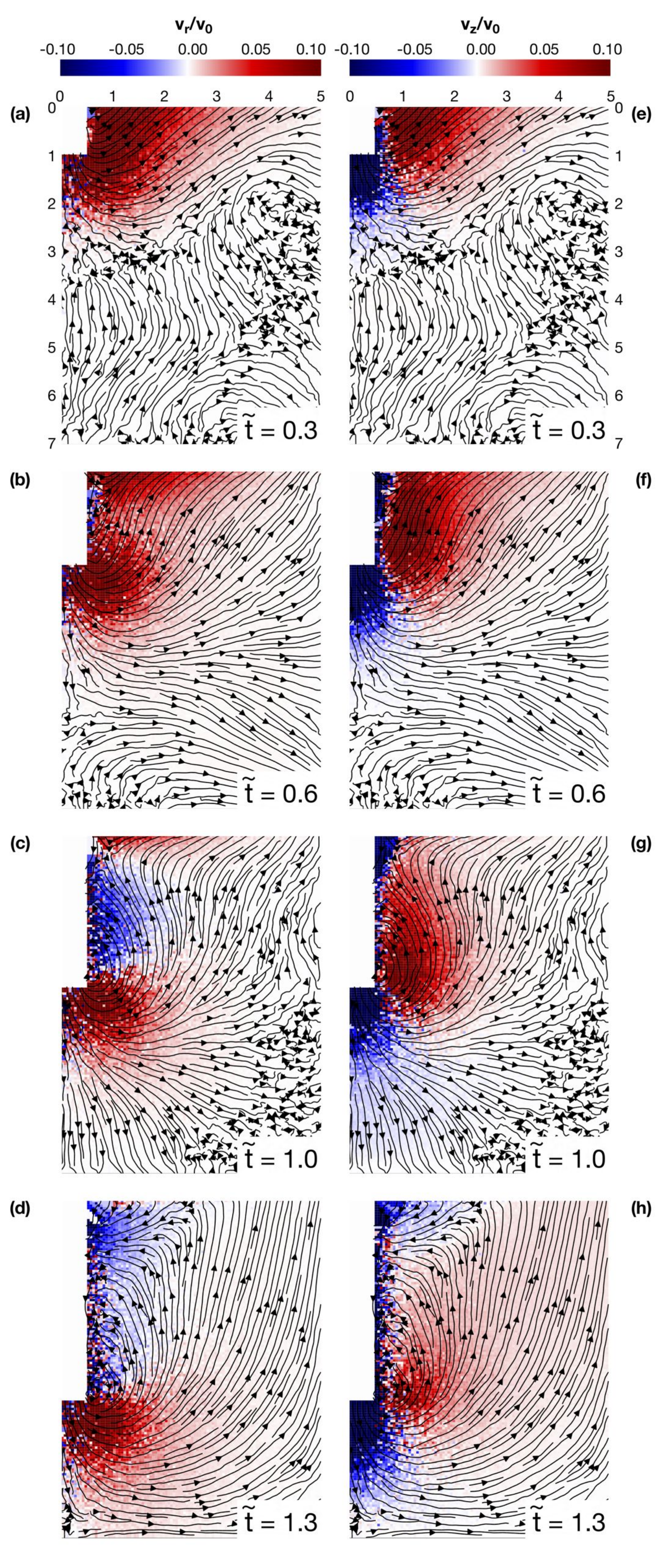}
\caption{\label{fig:normx_stream} Stream plots with vertical (\textit{right}) and radial (\textit{left}) velocity fields for $v_0 = 140$ cm/s at scaled times before and after $\widetilde{t}=1$.}
\end{figure}

Because the granular dynamics below the rod surface appear to be responding to the force transition, rather than causing it, we now investigate the flow characteristics between the surface of the rod and the surface of the bed.
Fig.~\ref{fig:normx_stream} shows the radial (\textit{a} - \textit{d}) and vertical (\textit{e} - \textit{h}) components of the granular velocity at four different scaled times.
At early times (Fig.~\ref{fig:normx_stream}, \textit{a} and \textit{e}), the flow above the intruder surface is uniformly upward and outward, excepting the grains entrained near the shaft of the rod.
This is still largely true at $\widetilde{t}=0.6$ (\textit{b} and \textit{f}), though the radial velocity intensity is becoming increasingly weak between the area of high shear near the rod surface and the free boundary of the bed.
At the critical time (\textit{c} and \textit{g}), the flow has qualitatively changed: the flow lines now fold back radially into the rod, while the vertical velocity of a wedge of grains at the rod surface has become negative.
This trend continues after $\widetilde{t}=1$ (\textit{d} and \textit{h}), as the flow lines originating at the rod surface no longer have a direct route to the surface.
The grains near the rod continue to fall downwards and inwards, while the area of high velocity that was concentrated near the shaft of the rod at early times is now obliged to fan out into the bulk and the flow lines expand to cover nearly the entire system.

\begin{figure}[h]
\includegraphics[width=\columnwidth]{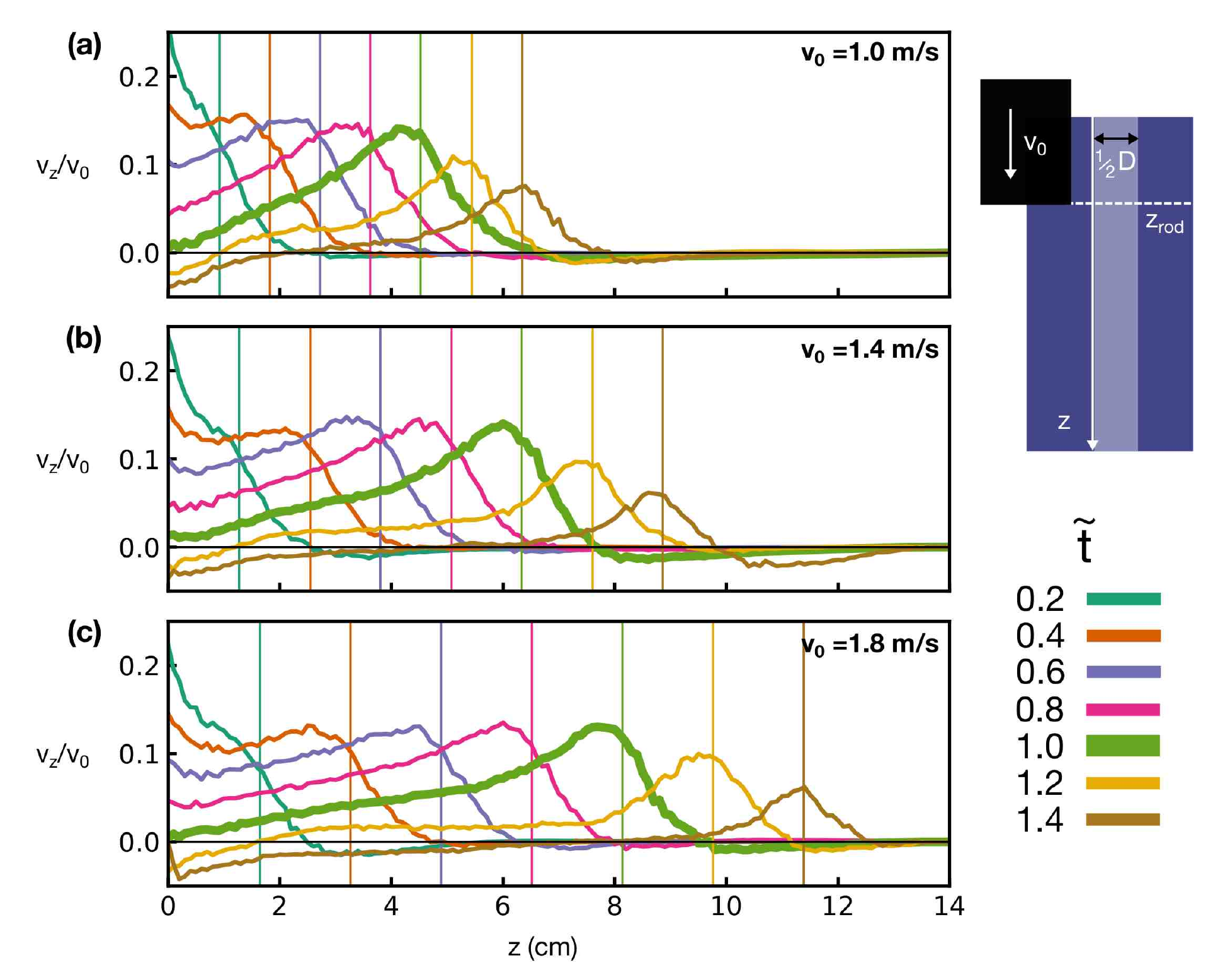}
\caption{\label{fig:vz_profile} Vertical velocity profile in a strip of grains from 0.5 cm to 1.5 cm of the rod shaft is plotted as a function of depth with respect to the bed surface (diagrammed at \textit{right}). Parallel vertical lines indicate the position of the rod in the bed. The color of the lines corresponds to the scaled time $\widetilde{t}$ for each $v_0$, and the bold line denotes $\widetilde{t}=1$. A positive value of $v_z$ means that grains are moving upwards, opposite to the velocity of the rod. \textit{a)} $v_0 = 1.0$ m/s. \textit{b)} $v_0 = 1.4$ m/s. \textit{c)} $v_0 = 1.8$ m/s. }
\end{figure}

To make this evaluation more quantitative, we inspect the vertical velocity profile of the grains near the rod shaft as a function of time for three different impact speeds.
In Figure~\ref{fig:vz_profile}, the vertical velocity profile of a strip of grains near the rod shaft (see Fig.~\ref{fig:vz_profile}, \textit{right}) for three impact speeds is plotted in a stationary frame relative to the surface of the bed.
The instantaneous position of the rod is indicated with vertical lines, and the profile corresponding to the critical time is shown in bold.
The same trends identified in the flow fields are present here.
While $\widetilde{t}<1$, the vertical velocity is positive from the surface of the bed down to the rod position, indicating a positive mass flux as grains are funneled upwards.
As $\widetilde{t}$ approaches 1, the vertical velocity becomes lopsided: there is still a high volume of grains being displaced by the rod and forced upwards, but their path to the surface is increasingly choked off.
Finally, at $\widetilde{t}=1$, the surface of the bed reaches $v_z = 0$ for all $v_0$, and after this point will continue to accumulate negative velocity.
Though $\widetilde{t}=1$ happens at different rod penetration depths in the bed, the evolution of the flow field is governed by time.

\begin{figure}[h]
\includegraphics[width=\columnwidth]{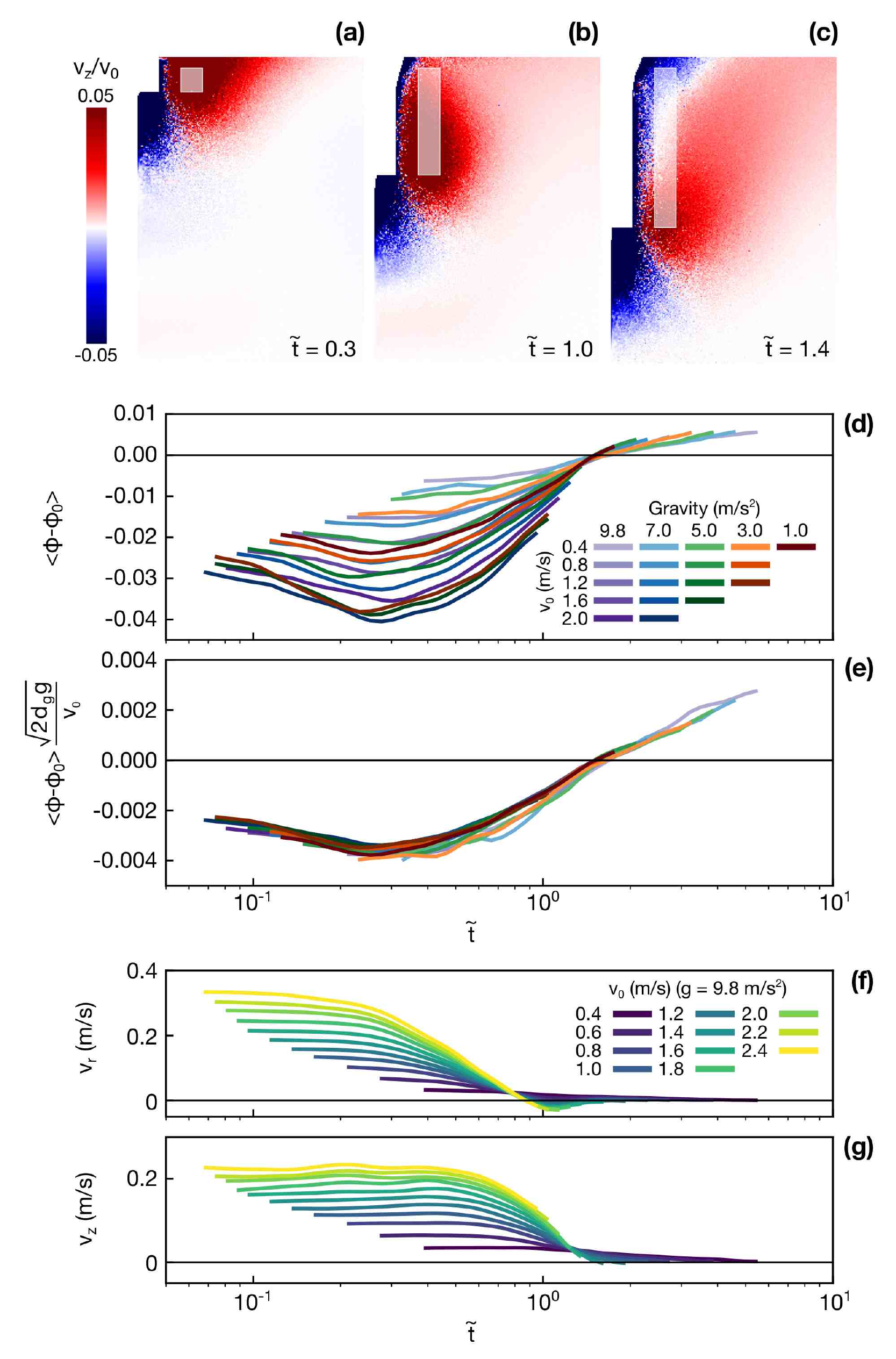}
\caption{\label{fig:pf_vz_avg_23} \textit{a)-c)} Heatmap of the vertical component of grain velocity, $v_z$, for intruder speed $v_0 = 120$ cm/s, at three different values of $\widetilde{t}$. A qualitative change in flow behavior can be observed as $\widetilde{t}$ exceeds 1. The vertical annulus of grains considered in the following subplots is highlighted in white: from $r=2$ to $r=3$ cm, and from $z=0.5$ cm to $z=z_{rod}$. \textit{d)} Average change in packing fraction in the segment of grains near the rod shaft shaded in \textit{a}-\textit{c}. A negative value indicates a decrease in packing fraction. \textit{e)} The change in packing fraction, scaled by the ratio $v_0/v_c$. \textit{f)} Average radial velocity, $v_r$, as well as the average vertical velocity, $v_z$, (\textit{g}) in this segment. The radial velocity decreases below zero, indicating that grains are moving back towards the rod as $\widetilde{t}$ nears 1.}
\end{figure}

Another perspective is shown in Fig.~\ref{fig:pf_vz_avg_23}, helping to make the time dependence of flow characteristics more explicit.
We average the grain velocity and packing fraction in a strip of grains of width $R$ (shown as the shaded region in Fig.~\ref{fig:pf_vz_avg_23}, \textit{a}-\textit{c}), in order to avoid the effect of grain entrainment at the shaft surface.
The effective annulus of grains over which the flow quantities are averaged reaches from just under the surface of the bed to the surface of the rod, so that it is growing in time.
If we assume that there is some degree of feedback between the force felt by the rod and the route taken by the excavated grains in their journey to the surface, the averaged flow quantities in this region provide a snapshot of the amenability of these grains to convection.
In Fig.~\ref{fig:pf_vz_avg_23}, \textit{d}, the average change in packing fraction is plotted for different values of gravity.
During the velocity dependent force regime, this segment is indeed fluidized: the average packing fraction in this region decreases by as much as 4 percentage points, where, unsurprisingly, the highest $v_0$ corresponds to the highest decrease in $\phi$.
In Fig.~\ref{fig:pf_vz_avg_23}, \textit{e}, we are able to collapse the change in packing fraction with $v_0/v_c$.
Though not constant in time, the relative fluidization of the bed appears to be directly proportional to the ratio of the impact speed to the gravitational settling velocity of individual grains.

Finally, the average of the radial (Fig.~\ref{fig:pf_vz_avg_23}, \textit{f}) and vertical (\textit{g}) velocities follow the trend observed qualitatively in Fig.~\ref{fig:normx_stream}.
Both the average radial and vertical velocities indicate grain motion up and away from the rod for $\widetilde{t}<1$.
At very early $\widetilde{t}$, the grain velocity is almost independent of time, but begins to decay once $\phi-\phi_0$ reaches its minimum value.
Near $\widetilde{t}=1$, the radial velocity briefly becomes negative, corresponding to the inward flow in Fig.~\ref{fig:normx_stream}, \textit{c}, but at this point both the radial and vertical average velocities quickly deteriorate.

\begin{figure}[h]
\includegraphics[width=\columnwidth]{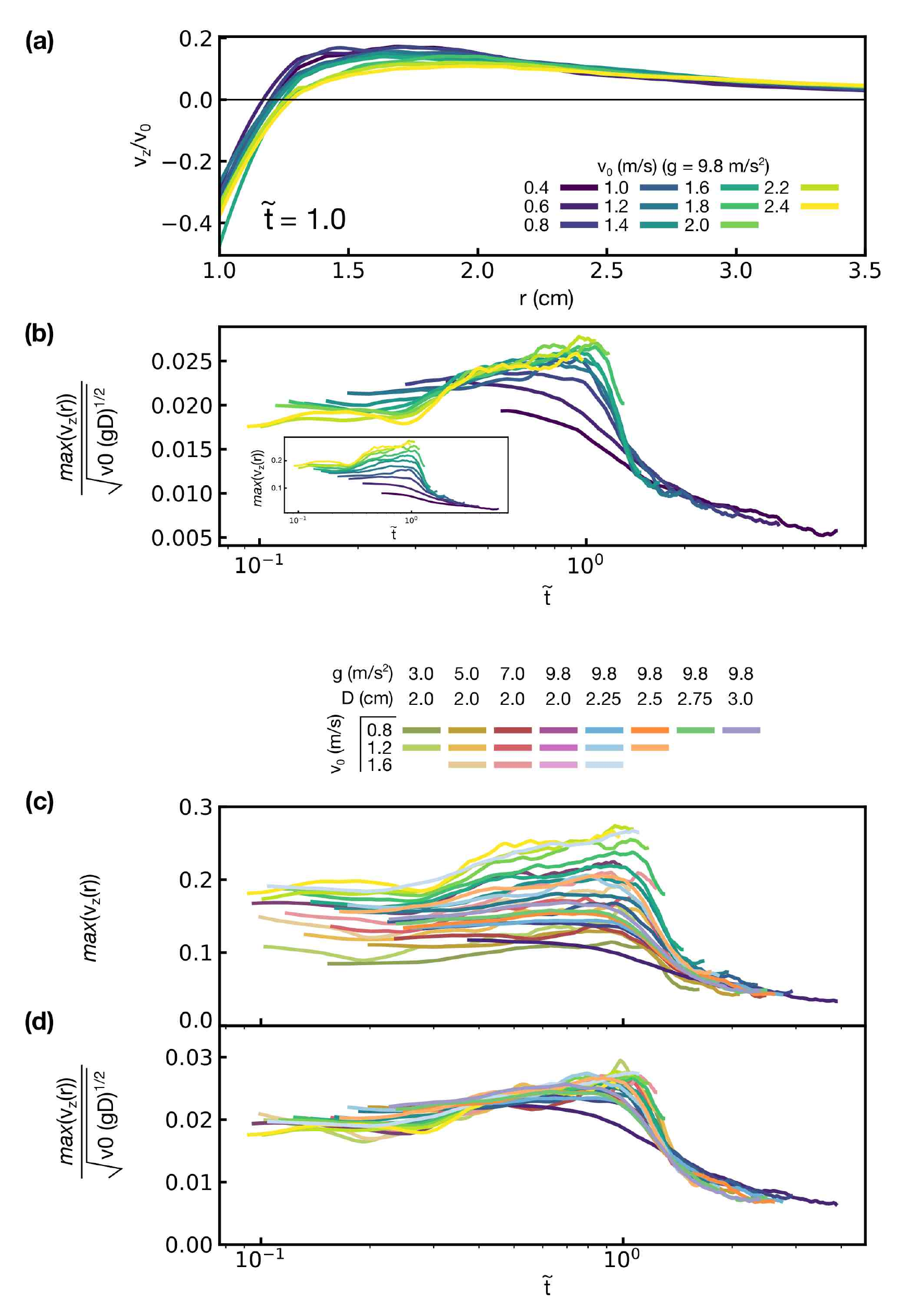}
\caption{\label{fig:min_Vz} \textit{a)} Vertical velocity profile of width $2r_g$ taken radially from the rod surface for $R=1$. Near the rod, at $r \rightarrow R$, grains are entrained in the motion of the rod and move downwards. Away from the shaft, the grain velocity $v_z$ becomes positive, indicating an upwelling of grains, and then decays as it travels in the bulk. \textit{b)} Maximum vertical grain velocity in the profiles shown in \textit{a}, normalized by a factor proportional to $v_0$ and the characteristic velocity, $\sqrt{gD}$ (unnormalized in \textit{inset}). The lowest speed does not collapse as well as $v_0>40$ cm/s, indicating a minimum fluidization speed. \textit{c)} and \textit{d)} The same collapse for different values of $D$, $g$ and $v_0$, for $v_0 > 40$ cm/s. The sharp decrease in maximum vertical grain velocity at $\widetilde{t}=1$ correlates with the activity of the uppermost layer of grains.}
\end{figure}

Fig.~\ref{fig:min_Vz} relates the time dependent transition identified in Fig.~\ref{fig:pf_vz_avg_23} for grains above the rod position, to the characteristics of the vertical velocity of grains parallel with the rod surface.
In Fig.~\ref{fig:min_Vz}, the vertical velocity profile in a plane parallel with the rod surface is plotted as a function of $r$.
The vertical velocity reaches a peak $\sim 1.5-2.0 R$ from the rod shaft, and then decays into the bulk.

Plotting the maximum vertical velocity in this plane as a function of $\widetilde{t}$ (Fig.~\ref{fig:min_Vz}, \textit{b}) mirrors, to a certain extent, the packing fraction collapse shown in Fig.~\ref{fig:pf_vz_avg_23}, \textit{e}.
By normalizing the maximum upward velocity at the rod surface --- which may be assumed to characterize the strength of the flow field in this plane, given the similar profile structures in Fig.~\ref{fig:gam_vz_pf_prof}, \textit{a} --- by the product of $v_0$ and the gravity-scaled velocity $\sqrt{gD}$, we are able to achieve a serviceable collapse that is particularly successful at the faster impact speeds tested.
Although not immediately clear, the quantity $\sqrt{v_0\left( gD \right)^{1/2}}$ could represent the geometric mean of the two principal velocity scales in the system at this point of critical juncture, a mixing of the two spatial regimes: the area directly below the intruder, dominated by the impact speed $v_0$, and the volume of grains above the rod surface that is governed by the gravitational velocity scale $\sqrt{gD}$.
Nonetheless, the maximum upward velocity, and with it the magnitude of the entire velocity profile at the rod surface, decreases dramatically at $\widetilde{t}=1$.
Because the transition from high to low upward velocity is much sharper and less foreshadowed than the vertical velocity crossover at $\widetilde{t}=1$ in Fig.~\ref{fig:vz_profile} or the averaged velocities in Fig.~\ref{fig:pf_vz_avg_23}, it could be that the vertical velocity profile of the grains parallel with the rod surface is responding to the conditions imposed by the grains above --- and not causing the transition itself.

\begin{figure}[h]
\includegraphics[width=\columnwidth]{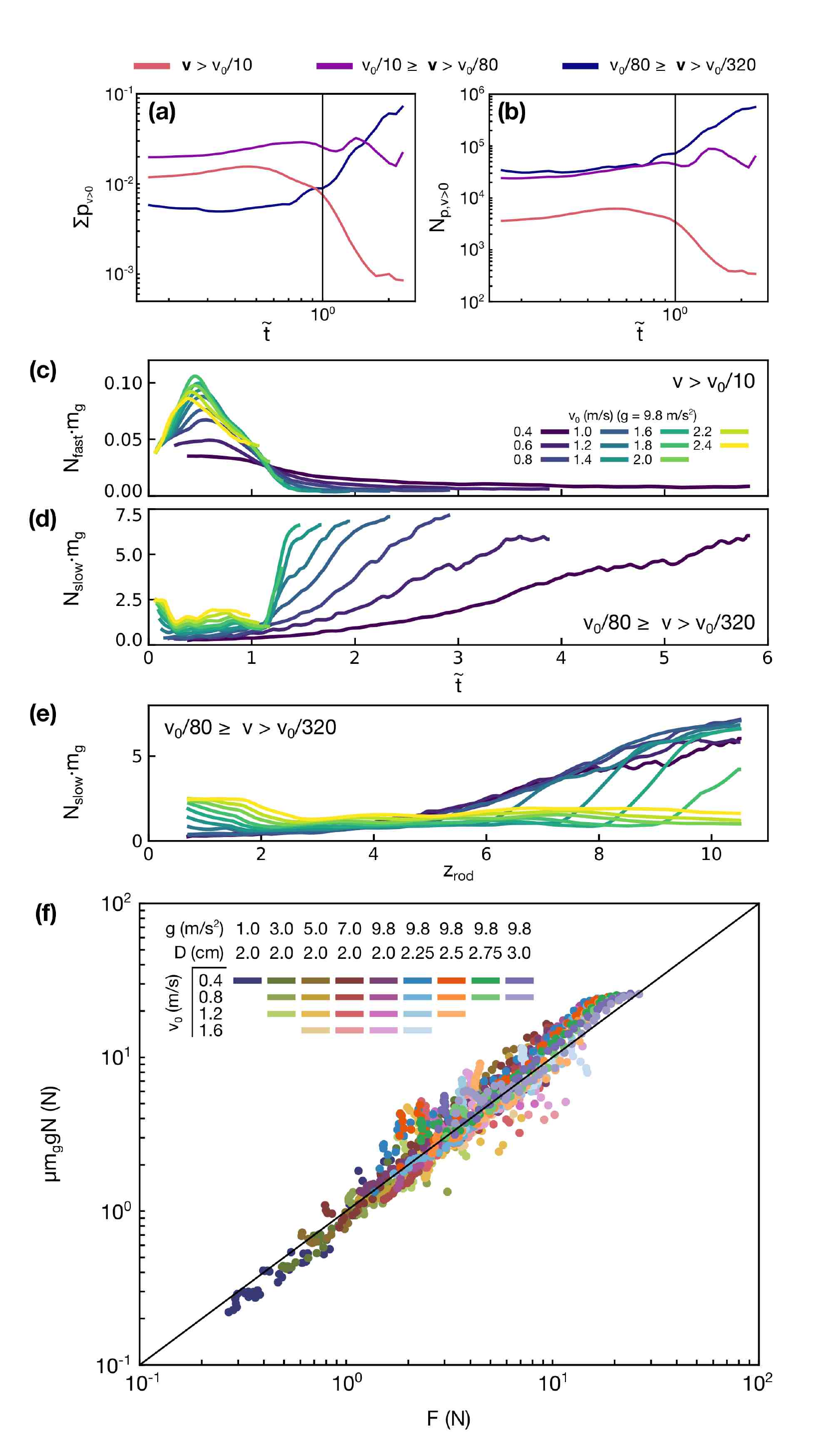}
\caption{\label{fig:mom_count} \textit{a)} Summed upward momentum and bin counts (\textit{b}) of all grains with $r > R$, binned by grain velocity, for $v_0 = 100$ cm/s. When $\widetilde{t} < 1$, the majority of the granular momentum is contained in relatively high velocity grains. \textit{c)} Summed momentum of the high velocity grains, defined as grains with $v > v_0/10$, as a function of dimensionless time. \textit{d)} Summed upward momentum of low velocity grains, defined as grains with $v_0/320 < v \leq v_0/80$. \textit{e)} Summed upward momentum of low velocity grains plotted as a function of rod position. \textit{f)} Drag force on the rod plotted against the summed mass of all grains with $v_z > v_0/320$, multiplied by $\mu g$. A 1:1 line is plotted for reference.}
\end{figure}

Both the vertical velocity profile at the rod surface (Fig.~\ref{fig:min_Vz}, \textit{d}), and the mean velocity of the grains above the rod (Fig.~\ref{fig:pf_vz_avg_23}, \textit{f} and \textit{g}), decline sharply in magnitude at $\widetilde{t}=1$, at the same time as the drag force $F$ assumes the quasistatic limit.
To make sense of this, we will now look at the system through the lens of the momentum carried by the grains over the course of the impact.
As we noted in Fig.~\ref{fig:normx_stream}, as $\widetilde{t}$ increases beyond $1$, the stream lines transition from involving chiefly the grains near the rod, to expanding across the entire system.
When $\widetilde{t}<1$, the upward velocity of the grains being displaced by the rod is high (Fig.~\ref{fig:min_Vz}, \textit{d}), and, as shown in Fig.~\ref{fig:pf_vz_avg_23}, \textit{e}-\textit{g}, a quantity of grains near the rod shaft is both flowing upwards and at least partially fluidized, providing an easy route to the surface for displaced grains.

As $\widetilde{t}$ increases beyond 1.0, the mass flux of grains at the surface of the bed near the rod passes through zero and becomes negative (Fig.~\ref{fig:vz_profile}), effectively confining the flow field created by the rod as it penetrates into the bed (see Fig.~\ref{fig:heatmaps}).
In the absence of a fluidized path to the surface, the displaced grains begin to transfer their momentum to the bulk, which creates the low velocity but wide reaching stream lines in Fig.~\ref{fig:normx_stream}, \textit{h}.
Here, we will evaluate the amount of momentum carried by grains of different instantaneous velocity magnitudes in the system.
The `fast' grains are defined as those whose upward velocity is at least $v_0/10$, the intermediate grains as those whose upward velocity lies between $v_0/10$ and $v_0/80$, and the slowest grains carry an upward velocity from $v_0/80$ down to $v_0/320$.
Below this velocity it is increasingly difficult to disentangle system-wide noise from grains that are truly participating in the flow, and in any case the number of grains with $v_0/320 > v_z \geq v_0/640$ is independent of $v_0$ and time.

In Fig.~\ref{fig:mom_count}, \textit{a} and \textit{b}, the net momentum in the system is summed as a function of $\widetilde{t}$, as well as the number of grains occupying each momentum bin, for an intruder speed of $v_0=1.0$ m/s.
When $\widetilde{t}<1$, the momentum carried by relatively fast grains is much higher than that carried by the slowest grains, despite there being more slow grains participating (Fig.~\ref{fig:mom_count}, \textit{b}).
However, after $\widetilde{t}=1$, the momentum carried by fast grains drops precipitously, while that carried by slow grains grows over the same period.
Additionally, the absolute number of grains carrying a low amount of momentum begins to grow as the number of high velocity grains dwindles.

In Fig.~\ref{fig:mom_count}, \textit{c} and \textit{d}, the mass of grains in possession of a `fast' or `slow' upward velocity is plotted for a range of $v_0$.
At early times, the mass of fast grains peaks before decaying and levelling off at a negligible value after $\widetilde{t}=1$.
In contrast, the much greater mass of slow grains involved in the flow begins a meteoric increase at the critical time.
Plotting the mass of slow grains as a function of rod position can help make the relationship between the number of low speed grains engaged in the flow (Fig.~\ref{fig:mom_count}, \textit{e}) and the drag force (Fig.~\ref{fig:force_scaling}, \textit{a}) more apparent.

At depths below $z_{rod}<3$, the slight increase in mass is likely due to the propagation of compression fronts throughout the system that were created during yielding.
After this deviation, the mass of slow velocity grains ($M_{slow}(v_0,z)$) remains steady for the highest impact speeds, though the overall constant depends on $v_0$.
For force traces that reach $\widetilde{t}=1$, the mass of slow velocity grains jumps up to join a linearly increasing curve shared by the slowest $v_0$.
This carries two implications.
Firstly, in contrast with the drag force, $M_{slow}$ for higher $v_0$ does not join the low velocity limit when the curves intersect.
This suggests again that the involvement of increasing numbers of low velocity grains is a consequence of the external flow constraints, and not a cause in and of itself.
Indeed, at the $z_{rod}$ corresponding to $\widetilde{t}=1$, $M_{slow}$ undergoes a rapid rise from its initial, constant value, to catch up with the linear quasistatic curve.
Secondly, the linear increase of $M_{slow}$ with depth suggests that the drag force $F$ is directly related to this quantity.

In Fig.~\ref{fig:mom_count}, \textit{f}, we plot the net normal force of the mass of grains moving upward with a velocity of at least $v_0/320$, multiplied by the friction coefficient $\mu$, against the net drag force felt by the intruder.
Though there is scatter, the drag force on the rod, for all impact speeds, strengths of gravity, and rod diameters tested, is consistent with the frictional force required to move an increasing number of low velocity grains.
This conclusion partially supports that arrived at by Kang \textit{et al.}, which predicted that the disproportionate strength of the linear depth dependence is due to the involvement of a broad volume of grains that are able to interact with the intruder~\cite{blumenfeld_nature}.

However, as stated earlier, the involvement in the flow of increasing numbers of low velocity grains appears to be an obligation due to the confinement of the flow field, rather than an immutable aspect of impact that is necessarily present at all velocities or a separate force that acts independently.
This would imply that, in at least the case of constant velocity impact, the depth dependent force itself is a symptom of confined flow fields imposed by the settling action of grains above the surface of the intruder --- a distortion of the velocity dependent drag force under certain empirical conditions.

\section{\label{sec:discussion}Discussion}

We have performed constant speed penetration experiments at impact speeds ranging from the quasistatic to the inertial, and have discovered a surprising transition between velocity dependent and velocity independent behavior that is not predicted by existing granular drag models (Fig.~\ref{fig:exp_slow_fast}, \textit{b} - \textit{c}).
In studying this effect further, we performed calibrated, large-scale MD simulations that enabled a direct investigation of the mechanisms for the observed velocity independent behavior.

We observed a strong initial peak, proportional to $v_0^2$, corresponding to the yield force (Fig.~\ref{fig:peak_pf})~\cite{goldman_force_flow_pre}.
The shape and magnitude of this force is likely sensitive to the shape of the intruder tip.
Because, in our system, $\phi_0>\phi_c$, as the grains begin to flow they must dilate in a region under the rod surface, producing an area of low density surrounding the rod (Fig.~\ref{fig:heatmaps})~\cite{goldman_force_flow_pre}.

A velocity dependent drag force follows the onset of flow, whose extent depends on a time scale $\widetilde{t}=(z/v_0)\sqrt{g/D}$(Fig.~\ref{fig:force_scaling}).
When $\widetilde{t}=1$, the drag force experienced by the rod becomes linearly dependent on depth, connecting with and following the quasistatic drag limit.

The crossover from a velocity to a depth dependent drag force is caused by the fluidization --- and subsequent settling --- of the volume of grains above the rod surface (Fig.~\ref{fig:normx_stream} and Fig.~\ref{fig:vz_profile}).
At early times, if $v_0 > v_c$, the grains within a few $D$ of the rod shaft are dilated and kicked upwards, creating a region of low packing fraction and high velocity that funnels displaced grains from the rod face to the surface of the bed (Fig.~\ref{fig:pf_vz_avg_23}).

However, at $\widetilde{t}=1$, the fluidized grains begin falling back towards the bottom of the box, isolating the flow around the rod from an easy route to the surface of the bed.
The confined flow field is forced to transfer momentum and displaced grains throughout the bulk of the bed, which can be seen in the transition in the velocity of the main carriers of momentum (Fig.~\ref{fig:min_Vz} and Fig.~\ref{fig:mom_count}).
We find that the drag force is approximately equal to the total number of grains being forced to flow upwards, consistent with the prediction of Kang \textit{et al.}~\cite{blumenfeld_nature}.

When $v_0$ increases beyond the quasistatic regime, the resulting dynamics are unexpected, yet rich.
The actions of the grains above the position of the intruder determine whether the drag force is inertial, corresponding to localized mass flux and a high degree of fluidization near the rod shaft, or quasistatic, in which the flow field around the rod face is insulated from the surface and forced to enter dialogue with the bulk (see Fig.~\ref{fig:vz_profile} and Fig.~\ref{fig:mom_count}).
Though this effect is robust under conditons of constant velocity, it is not clear how the motions of the ejected grains would affect the drag experienced during projectile penetration.
Additionally, the area of dilation surrounding the rod may have an effect on the drag force under conditions of rapid deceleration.
Further experiments and simulations would be required to evaluate the transference of these observations to projectile impact.
Finally, the role of friction is unclear in this framework, especially as the frictionless limit is approached.
Further work would be necessary to understand how the flow characteristics and the rescaled time are affected by friction.

\begin{acknowledgments}
I extend sincere gratitude to my advisor, Heinrich M. Jaeger, for his support.
I thank Kieran Murphy, Melody Lim, Abhinendra Singh, and Endao Han for their invaluable discussions, and I am very grateful to Paul Umbanhowar for the generous use of his glass beads throughout this project.

This work was supported by the Center for Hierarchical Materials Design (CHiMaD), which is supported by the National Institute of Standards and Technology, US Department of Commerce, under financial assistance award 70NANB14H012, and by the Army Research
Office under Grant Number W911NF-19-1-0245.
\end{acknowledgments}







\providecommand{\noopsort}[1]{}\providecommand{\singleletter}[1]{#1}%

\end{document}